%% file: CCmodel_Letter_final-Heitor.tex
\documentclass[reprint,superscriptaddress,footinbib,nobibnotes,eprint,amsmath,amssymb,aps,prb,citeautoscript,final,floats]{revtex4-2}

\usepackage[utf8]{inputenc}
\usepackage{latexsym}
\usepackage{amsmath}
\usepackage{amssymb}
\usepackage{amsfonts}
\usepackage{graphicx}
\graphicspath{ {graphs/} }
\usepackage{slashed}
\usepackage{xcolor}
\usepackage{soul}
\usepackage{braket}
\usepackage{MnSymbol}
\usepackage{wasysym}
\usepackage{mathrsfs}  
\usepackage{subcaption}
\usepackage{physics}
\usepackage{changes}
\usepackage{verbatim}
\usepackage{microtype}
\usepackage{color}

\usepackage[colorlinks=true]{hyperref}

\makeatletter
\def\maketitle{
	\@author@finish
	\title@column\titleblock@produce
	\suppressfloats[t]}
\makeatother

\begin{document}
	
	\title{Fractal Subsystem Symmetries, 't Hooft Anomalies, and UV/IR Mixing}
	
	\author{Heitor Casasola}
	\email{heitor.casasola@uel.br}
	\affiliation{Departamento de F\'isica, Universidade Estadual de Londrina, 86057-970, Londrina, Paraná, Brazil}
	\affiliation{Physics Department, Boston University, Boston, Massachusetts, 02215, USA}
	
	\author{Guilherme Delfino}
		\email{delfino@bu.edu}
	\affiliation{Physics Department, Boston University, Boston, Massachusetts, 02215, USA}
		
	\author{Pedro R. S. Gomes}
	\email{pedrogomes@uel.br}
	\affiliation{Departamento de F\'isica, Universidade Estadual de Londrina, 86057-970, Londrina, Paraná, Brazil}
	
	\author{Paula F. Bienzobaz}
	\email{paulabienzobaz@uel.br}
	\affiliation{Departamento de F\'isica, Universidade Estadual de Londrina, 86057-970, Londrina, Paraná, Brazil}
	
	\received{8 November 2023}
	\revised{31 January 2024}
	\accepted{7 February 2024}
	\published{27 February 2024}

\begin{abstract}	
		
In this work, we study unconventional anisotropic topologically ordered phases in $3d$ that manifest type-II fractonic physics along submanifolds. While they behave as usual  topological order along a preferred spatial direction, their physics along perpendicular planes is dictated by the presence of fractal subsystem symmetries, completely restricting the mobility of anyonic excitations and their bound states. We consider an explicit lattice model realization of such phases and proceed to study their properties under periodic boundary conditions and, later, in the presence of boundaries. We find that for specific lattice sizes, the system possesses line and fractal membrane symmetries that are mutually anomalous, resulting in a nontrivially gapped ground state space. This amounts to the spontaneous breaking of  the fractal symmetries, implying a subextensive ground state degeneracy. For the remaining system sizes the fractal  symmetries are explicitly broken by the periodic boundary conditions, which is intrinsically related to the uniqueness of the ground state. Despite that, the system is still topologically ordered since locally created quasiparticles have nontrivial mutual statistics and, in the presence of boundaries, it still presents anomalous edge modes. The intricate symmetry interplay dictated by the lattice size is a wild manifestation of ultraviolet/infrared (UV/IR) mixing. 
\\

\noindent DOI: \href{https://doi.org/10.1103/PhysRevB.109.075164}{10.1103/PhysRevB.109.075164}

\end{abstract}

\maketitle
	
\tableofcontents
	

\section{Introduction}

It is quite remarkable that certain elementary systems, containing simple degrees of freedom that  interact only locally, can give rise to exotic forms of matter \cite{Wen_2017}. Topologically ordered systems are among the most prominent examples, with emergent quasiparticle excitations carrying anyonic statistics. Fractonic topological ordered systems are even more exotic \cite{Chamon_2005,Bravyi_2011,Haah_2011,Vijay:2015mka,Nandkishore2017,SlagleKim2017,You:2018zhj,Shirley2019,Chen2019a,Fuji2019,Fontana:2020tby,Fontana:2021zwt,Nandkishore:2018sel,Pretko:2020cko}, with excitations that are intrinsically devoid of mobility due to the existence of generalized forms of symmetries known as {\it subsystem} symmetries \cite{Vijay:2016phm, Pretko2017,Jangjan2022,Delfino2023}.
	
Subsystem symmetries are generated by conserved charges along rigid spatial subdimensional manifolds, which tend to be extremely sensible to the underlying lattice geometry. Compliance with such conservation laws imposes severe restrictions on the mobility of the excitations. This contrasts with usual topological order, which is in general characterized by topologically deformable symmetry generators, known as {\it higher-form} symmetries, and poses no constrains on excitation's mobility \cite{Gaiotto:2014kfa,Gomes:2023ahz,Brennan:2023mmt,Luo:2023ive,Bhardwaj:2023kri}.

Abelian topological order and fractonic phases may be rephrased in terms of spontaneous breaking of, respectively, finite \textit{higher-form} and \textit{subsystem symmetries}  \cite{,Wen_2019,Qi:2020jrf,Rayhaun:2021ocs}. This follows from the definition of topological order in terms of local indistinguishability of ground states \cite{Bonderson_2013,Huxford:2023bhb}. Additionally, any two ground states are connected by {\it extended} symmetry operators, which  precisely fits the notion of spontaneous symmetry breaking in the context of a generalized symmetry \cite{McGreevy_2023}. Generalized symmetries appear to distinguish themselves from usual symmetries, as they can be present in low-energy states even when the Hamiltonian is not symmetric \cite{McGreevy_2023,Pace_2023emergent, cherman2023emergent}. Furthermore, although in this work we only discuss systems containing Abelian anyons---usually associated to higher-form and subsystem symmetries---it is worth mentioning  that non-Abelian anyons can also be casted in the language of generalized \textit{non-invertible} symmetries \cite{Froehlich2004,FROeHLICH2010,Chang2019,SchaferNameki2023,Shao2023}.

An alternative way of understanding the nontriviality of the ground state space is through the nontrivial braiding statistics among excitations, which signals the presence of 't Hooft anomalies in Abelian topological order. The matching of anomalies from the ultraviolet (UV) to the infrared (IR) imply that the ground state must be nontrivially gapped \cite{tHooft:1979rat}.

It is common for $d$-dimensional systems invariant under subsystem symmetries, with support in dimensions smaller than $d$, to have a subextensive number of conserved charges. Typical examples are type-I fracton systems that, when defined on a $3d$ $L \times L \times L$ system, possess charge conservation laws in $O (L)$ individual planes \cite{Nandkishore:2018sel, Pretko:2020cko}. This leads to a macroscopically large amount of symmetry generators, which in turn implies an enormous degeneracy of states. In particular, the ground state degeneracy (IR feature) is sensitive to the number of sites of the system (UV feature)---a phenomenon called UV/IR mixing. 

The emergence of UV/IR mixing from simple bosonic local theories has challenged our understanding of  effective theories and renormalization group. This follows from the fact that the IR physics depends sensitively on the UV details of the theory, and has been a major point of investigation \cite{Zhou:2021wsv,Seiberg_2020, Fontana:2020tby, You_2020, Lake2021rg}. In the context of generalized symmetries, several exactly solvable models have shed light on the origin of such phenomenon. In the context of gapped dipole moments (and higher-multipole momenta)  conserving gauge theories, emergent higher-form symmetries obey twisted boundary conditions in order for holonomy operators to close onto themselves \cite{Oh_2022, Pace_2022, Delfino_2023, Watanabe_2023}. For subsystem symmetries, which are closely related with systems studied in the work, the UV/IR mixing emerges from the fact that the system possesses an increasing number of symmetries as the system size grows \cite{Bravyi_2011, Vijay:2016phm,Ma_2017}.

A more dramatic manifestation of UV/IR mixing shows up in the case of subsystem symmetries where the charges are supported in sub-manifolds of fractal dimensions \cite{Newman_1999,Castelnovo_2012,Yoshida:2013sqa,Bulmash:2018knk,Devakul_2019,Zhou:2021wsv,Myerson_Jain_2022,Sfairopoulos2023}. These symmetries are intimately related to type-II fracton physics \cite{Haah_2011}, where no excitations are allowed to move. In this work we study fractal symmetries with support on Sierpinski triangles, which have Hausdorff dimension $\log(3) / \log(2)$. As a consequence, the low-energy properties are extremely sensitive to the lattice details and typically there is no uniform dependence of the ground state degeneracy with the lattice size as in the case of type-I fractons.  Additionally, we find that the fractal symmetries are always broken.  Whether it is spontaneously or explicitly broken, depends on the lattice linear size $L$. 
	
	
\section{Model}

\begin{figure}
	\includegraphics[height=2.5cm]{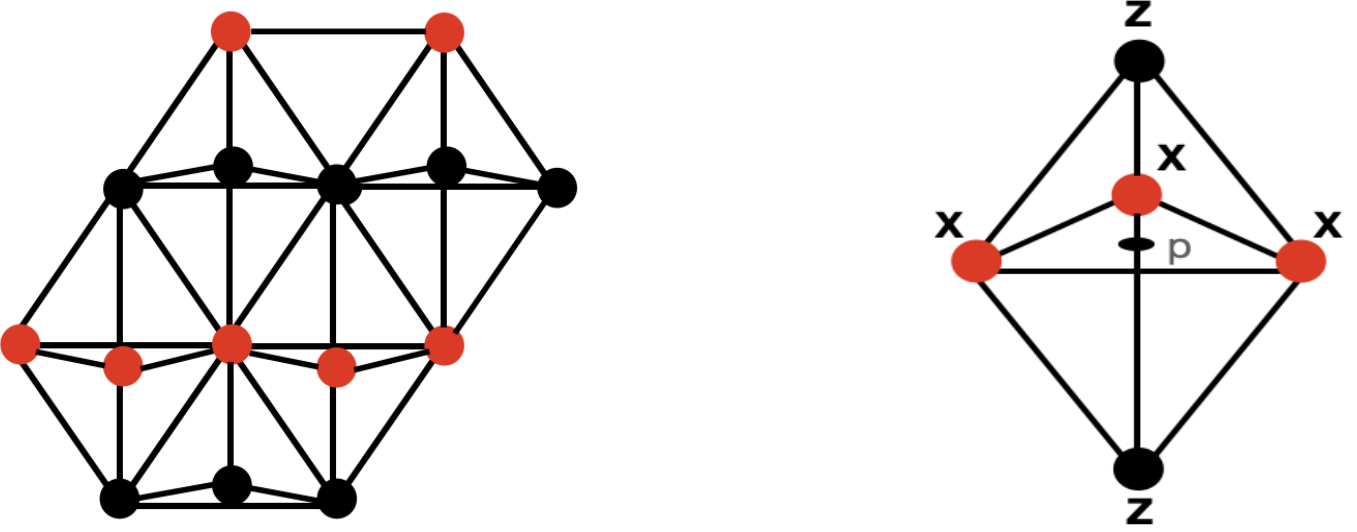}
	\caption{Three-dimensional hexagonal close-packed lattice and a TD operator.} \label{Lattice}
\end{figure}

The phases we are interested in are captured by the low-energy states of the fractal model introduced in \cite{Castelnovo_2012} in the context of topological quantum glassiness. The degrees of freedom of this model correspond to qubits located at the sites of a hexagonal close-packed lattice. The lattice is defined as the interpolating stack (along the $\hat z$ direction) of $d = 2$ triangular lattice planes that are dislocated in relation to each other by $a_0 ( \hat x/2 + \sqrt{3} \hat y/3 + \hat z/2 )$, where $a_0$ is the lattice spacing. We now focus on the lattice composed of centers $p$ of triangles, which can be decomposed into two sublattices $\Lambda = \Lambda_1 \oplus \Lambda_2$, as shown in Fig. \ref{Lattice}. The Hamiltonian is
\begin{equation} \label{0.1}
	H = - J \sum_{a = 1, 2} \, \sum_{p \, \in \, \Lambda_a} \mathcal{O}_p, \qquad J > 0,
\end{equation}
where the triangular dipyramidal (TD) operators $\mathcal{O}_p$ are given in terms of Pauli matrices $X$ and $Z$ according to
\begin{equation}
	\mathcal{O}_p \equiv Z_{p + \frac{1}{2} \hat{z}} \,
	X_{p + \frac{1}{2} \hat{x} - \frac{1}{2\sqrt{3}} \hat{y}} \,
	X_{p - \frac{1}{2} \hat{x} - \frac{1}{2 \sqrt{3}} \hat{y}} \,
	X_{p + \frac{1}{\sqrt{3}} \hat{y}} \,
	Z_{p - \frac{1}{2} \hat{z}},
\end{equation}
where $p$ is the center of the TD operator, as in Fig. \ref{Lattice}. 

A key property of the model \eqref{0.1} is that it is written in terms of commuting projectors, i.e., $\left[ \mathcal{O}_{p}, \mathcal{O}_{p^{\prime}} \right] = 0$ for any pair of operators, even when they share sites. Indeed, two neighboring TD operators can share one or two sites, and it is simple to see that in both cases they commute. 
Therefore, the energy spectrum is gapped and the energy levels are given by the sum of the eigenvalues of $\mathcal{O}_{p}$. As $\mathcal{O}_{p}^{2} = \openone$, the eigenvalues of $\mathcal{O}_{p}$ are  $\pm 1$ and ground states obey  $\mathcal{O}_{p} \ket{GS} = + 1 \ket{GS}$ for all $p$.


\section{Ground State Degeneracy}

Each lattice site hosts a two-dimensional Hilbert space, so that the specification of a state in a lattice with $N$ sites requires $2^N$ labels. As the total number of TD operators is equal to the number of lattice sites when the system is defined with periodic boundary conditions, it seems that the eigenvalues of the operators $\mathcal{O}_p$ are able to provide exactly the $2^N$ labels for the states. However, not all TD operators are independent. There are certain constraints that reduce the number of available labels and consequently increase the degeneracy. Let $N_c$ be the total number of constraints in the system. Then, the total number of available labels is $2^{N - N_c}$, so that the ground state degeneracy is $GSD = 2^N/2^{N - N_c} = 2^{N_c}$.

We can write the constraints as
\begin{equation} \label{eq:set1}
	\prod_{p \, \in \,\Lambda} \mathcal{O}_{p}^{t_{p}} = \openone,
\end{equation}
involving a set of weights $t_p$ mod 2. The dimension of the set  $\{ t_p \}$ gives the number of independent constraints $N_c$ of the model. 

\begin{figure}
	\includegraphics[height=2.5cm]{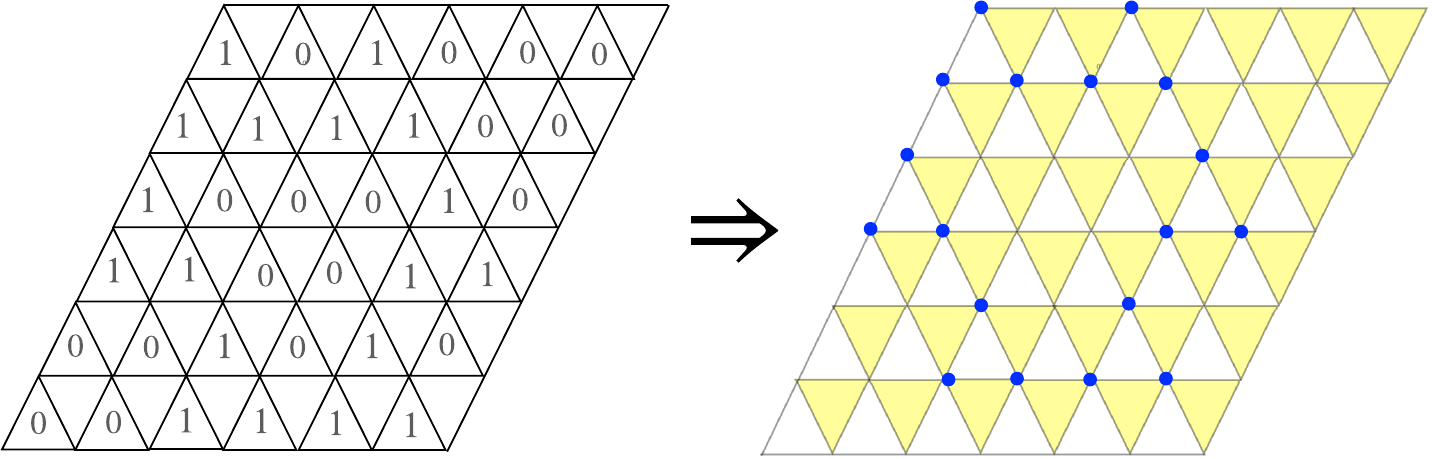}
	\caption{Membrane operator constructed from a constraint.} \label{constraint_membrane}
\end{figure}

Let us consider a periodic lattice with size $L \times L\times L_z$. For linear sizes $L = 2^n - 2^m$, with integers $n \ge 2$ and $0 \le m < n$, the number of constraints is $2^n - 2^{m + 1}$ for each one of the sublattices \footnote{See Supplemental Material \hyperref[SM]{Supplementary Material} for further insights and detailed explanations related to several topics discussed in the main text. These include the determination of \hyperref[MS1]{ground state degeneracy}, the \hyperref[MS3]{number of independent Wilson line operators}, how to obtain the \hyperref[MS4]{algebra among fractal membranes and Wilson line operators} and how\hyperref[MS5]{ Wilson lines can be simplified into a product of TD operators} for non-degenerate system sizes.}. Furthermore, the constraints produce fractal Sierpinski structures in the $xy$ planes. An example for $L=2^3-2^1=6$ is shown in the left of Fig. \ref{constraint_membrane}.

Lattice sizes of the form $L = 2^n - 2^m$ are not the only ones that possess nontrivial ground state degeneracy, as we can construct larger systems simply by taking multiple copies of the size $L$, namely, we can consider systems with sizes of the form $L' = k \left(2^n - 2^m\right) \neq  2^{n'} - 2^{m'}$. In this case, the ground state degeneracy is that of the building-block copy $L=2^n-2^m$, since it provides nontrivial solutions for $t$'s that are compatible with the periodic boundary conditions. As an example, take the size $L=9$. It is not of the form $2^n-2^m$, but it can be expressed as $L=3(2^2-2^0)$. The ground state degeneracy is the same as that one of the case $L=2^2-2^0=3$.

It may happen that there is more than one way to express certain size in terms of copies, i.e., $L = k \left( 2^n - 2^m \right) = k' ( 2^{n'} - 2^{m'} )$. In this case, we shall look for the building-block copy that provides the maximum number of nontrivial solutions for $t$'s, namely, which number among $2^n-2^{m+1}$ and $2^{n'}-2^{m'+1}$ is greater. This occurs for the smallest value of $k$ \cite{Note1}. Taking into account that there are two sublattices, we obtain the total number of constraints $N_c = 2 \left( 2^n - 2^{m + 1} \right)$. With this in mind, we can express in a unified way the ground state degeneracy associated with sizes
\begin{equation} \label{sizes}
L= k \left( 2^n - 2^m \right), \qquad 0 \le m < n - 1 \quad \text{and} \quad k \ge 1
\end{equation}
that is given by
\begin{equation} \label{gsd1}
	GSD = 2^{\left( 2^{n + 1} - 2^{m + 2} \right)},
\end{equation}
with $n$ and $m$ associated with the copy with the lowest value of $k$. Of course, whenever a size can be expressed with $k=1$, then it dictates the degeneracy. An example is $L=18$, which cannot be expressed with $k=1$. There are three ways to obtain $L=18$, namely, $3 (2^3-2^{1})$, $6 (2^2-1)$, and $9(2^2-2)$. According to the above discussion, the ground state degeneracy is given by \eqref{gsd1} with $n=3$ and $m=1$, which is constituted of three copies ($k=3$) of the size $L=6$.

The GSD in Eq. \eqref{gsd1} does not depend on $L_z$ due to the simple structure of the TD operators along the $z$ direction. Since it has two $Z$ Pauli matrices, one on top and one at the bottom, all the constraints involve  a product of TD operators along the whole $ z$ direction. This is analogous to usual topological constrains, as in the toric code \cite{Kitaev_2003}.	Despite of this, the third dimension is crucial to ensure topological ordering.

For system sizes $L \times L \times L_z$ in which $L$ cannot be expressed in the form \eqref{sizes}, the ground state is unique since there are no nontrivial solutions for $t$'s. We shall discuss these cases later. The intricate dependence of the ground state degeneracy on the size of the system is a severe manifestation of the UV/IR mixing. To stress this, consider $m = 0$ and a very large value of $n$. In this case, the system with the size $L = 2^{n} - 1$ has a huge degeneracy $GSD = 2^{\left( 2^{n + 1} - 2^{2} \right)}$, whereas the system with size just one unit larger $L = 2^n$ has a unique ground state.  

For general $L_x \neq L_y$ it is not clear the ground state degeneracy. However, for certain classes of system sizes, for example $L_x = q L_y$, for an integer $q$ and $L_x = 2^{n} - 2^{m}$, we can show that ground state degeneracy is $2^{2 \left(L_y -2^{m}\right)}$.


\section{Symmetry Operators and Mixed 't Hooft Anomalies}

A nontrivial ground state degeneracy is a {reflection} of a nontrivial algebra among certain symmetry operators, which indicates a nontrivial mutual statistics among anyons. For sizes of the form \eqref{sizes}, there are two types of subsystem symmetry operators: {\it fractal membrane} operators disposed in the $xy$ plane and {\it Wilson line} operators extended along the $z$ direction. 
	
Membrane operators are intimately related to the constraint structure \eqref{eq:set1} in the $xy$ plane (see relation (10) of the Supplemental Material \cite{Note1}). Consider $\prod_{z} \prod_{j \, \in \,\mathcal{M}_{I}^{a} (z)}\mathcal{O}_{j} = \openone$, where $j$ runs over all the sites belonging to $\mathcal{M}_I^a$ of the sublattice $a = 1, 2$ in the $xy$ plane, with $I = 1, \, \ldots \, , 2^n - 2^{m + 1}$ specifying the constraint. A subset of this product, ranging from $z = z_{1} + a_{0}/2$ to $z = z_{2} - a_{0}/2$, results in two membranes,
\begin{equation} \label{eq:membranes}
	\prod_{z = z_{1}}^{z_{2}} \, \prod_{j \, \in \,\mathcal{M}_{I}^{a}(z)} \mathcal{O}_{j} = M_I^b\left( z_{1} \right) \, M_I^b\left( z_{2} \right), \qquad b \neq a,
\end{equation}
where, $M_I^a$ is defined as
\begin{equation} \label{MO}
	M_I^a \equiv \prod_{j \, \in \, \mathcal{M}_I^a} Z_j.
\end{equation}
Each of these closed membrane operators individually commute with the Hamiltonian, since they contain either zero or two $Z$ operators acting on each TD operator (see Fig \ref{constraint_membrane}). Notably, while two membranes are obtained as a product of $\mathcal{O}_{j}$, an individual $M_{I}^{a}$ is a nontrivial symmetry\footnote{In analogy with the toric code, the two membranes of Eq. \eqref{eq:membranes} correspond to contractible loops and a single membrane corresponds to a non-contractible loop.}. As the constraints exhibit fractal Sierpinski structures in the $xy$ plane, the membrane operators resulting from this association will also enjoy such fractal pattern. It's worth mentioning that the membranes defined in Eq. \eqref{eq:membranes} are associated with the opposite sublattice of the constraint, as illustrated in Fig. \ref{constraint_membrane} by the blue dots acting on the yellow sublattice. 

\begin{figure}
	\includegraphics[width=5cm]{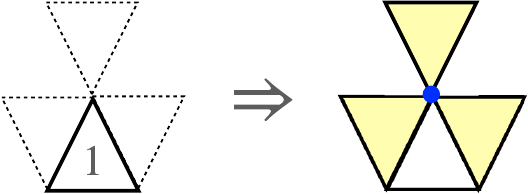}
	\caption{Association of an operator $\mathcal{O}$ partaking the constraint in the $xy$ plane in a sublattice with a $Z$ operator (blue dot) in the other sublattice (yellow).}  \label{rule}
\end{figure}

Consistent with Eq. \eqref{MO}, the membranes can be obtained by associating to each TD operator partaking the constraint and belonging to one of the sublattices a $Z$ operator acting in a site of the other sublattice, according to the map in Fig. \ref{rule}. This is reminiscent of the duality between dynamics and interactions \cite{Newman_1999}.

There is a one-to-one correspondence among the constraints, associated with nontrivial solutions for $\{t_p\}$, and the closed membrane operators. Accordingly, there is a total of $N_c = 2 (2^n - 2^{m + 1})$ linearly independent membrane operators. Although for every $z$-coordinate we can define a membrane operator as in Eq. \eqref{MO}, they are not regarded as independent. This is so because any two membranes $M_{I}^{a}(z_{1})$ and $M_{I}^{a} (z_{2})$, belonging to the same sublattice, and with same $I$, can be connected to each other through a product of TD operators. Thus, unless there is a defect, $\mathcal{O}_p \ket{\psi} = -\ket{\psi}$, between the planes at $z_1$ and $z_2$, the eigenvalues of $M_{I}^{a} (z_{1})$ and $M_{I}^{a} (z_{2})$ are constrained to be the same. Also, the membrane operators can be topologically deformed in the $z$ direction through local products of TD operators.

Wilson lines along the $z$ direction are constructed as $
W_{i}^a = \prod_{j\,\in\, \mathcal{L}_{i}^a} X_j,$
where $\mathcal{L}_{i}^a$ stands for a line along the $z$-axis crossing $xy$ planes at the site $i = (x,y)$, belonging to the sublattice $a$. If the line $\mathcal{L}_{i}^a$ is closed, the corresponding operator commutes with the Hamiltonian and then it is a symmetry operator. 

Wilson line operators are rigid and, in principle, for each sublattice, there are $L^2$ line operators, one for each $i = (x,y)$ in the $xy$ plane. However, not all of them are independent, as products of three lines can also be reduced to products of TD operators,  which act as the identity on the ground state. The number of independent Wilson lines is $2^n - 2^{m + 1}$ for each sublattice and that they are in one-to-one correspondence with the membrane operators \cite{Note1}. Accordingly, we label the Wilson lines with the same type of index $I$ as the membrane operators,
\begin{equation} \label{WL1}
	W_{I}^a = \prod_{j \, \in \, \mathcal{L}_{I}^a} X_j,
\end{equation}
where $\mathcal{L}_{I}^a$, with $I = 1, \, \ldots \, , 2^n - 2^{m + 1}$, corresponds to each one of the independent Wilson lines.  We can pair line and membrane operators such that a particular $W_I^a$ commutes with all other membrane operators except the one labeled by the same numbers $M^a_I$,
\begin{equation} \label{projective}
	M_I^a \, W_J^b = \left( -1 \right)^{\delta_{IJ}\delta_{ab}} \, W_J^b \, M_I^a,	
\end{equation}
with $a, b = 1, 2$ and $I, J = 1, \, \ldots \, , 2^n - 2^{m + 1}$ \cite{Note1}. This algebra leads to the ground state degeneracy $GSD = ( 2^{2^n - 2^{m + 1}} )^2$, reproducing \eqref{gsd1}. Such a nontrivial algebra among pairs of membrane and Wilson operators can be rephrased in terms of a mixed 't Hooft anomaly between the subsystem symmetries, preventing the ground state manifold to be trivially gapped.

We recall that a mixed 't Hooft anomaly is an obstruction to simultaneously gauging  the corresponding symmetries. This can be understood in an intuitive way. Consider a single pair of symmetry operators satisfying \eqref{projective} with $MW=-WM$, as well as states $\ket{\psi'}$ and $\ket{\psi''}$ constructed as
\begin{equation}
\ket{\psi'}= MW \ket{\psi}~~~\text{and}~~~ \ket{\psi''}= W M \ket{\psi}.
\end{equation} 
The commutation relation between $M$ and $W$ implies $ \ket{\psi''}=-\ket{\psi'}$. As the two states differ by a phase, they belong to the same ray and are identified actually as the same state. Now, if we try to gauge both symmetries, which means that the physical states must be invariant under the action of $M$ and $W$, we get a contradiction $\ket{\psi}=- \ket{\psi}$. In other words, there are no physical states in the gauged theory and the partition function vanishes.


\section{Excitations and Mutual Statistics}

Excitations correspond to any state with $\mathcal{O}_p \ket{\psi} = - \ket{\psi}$, for some TD operators. Such configurations are either created in pairs at the endpoints of open Wilson lines or in groups of three at the corners of open fractal membranes. The rigidity of string and membrane operators is reflected in the allowed dynamics for the quasiparticles excitations. While open lines $W_I^a$ are associated with the free transport of excitations along the $z$ direction, open $M_I^a$ cannot be used to move particles in the $xy$ plane. This is because $M_I^a$ maps a single particle state into a two-particle state, a process that has a high energy cost $\Delta E \sim J$. Open membranes $M_I^a$ can be used, though, to map three-particle states into three-particle states with no energetic cost, through the growth of the underlying fractal membrane $\mathcal M_I^a$. However, such a process passes through highly energetic intermediate states, making the corresponding tunneling probability very small. This implies that all excitations are immobile along the $xy$ plane in finite times, which is a manifestation of type-II fracton physics \cite{Haah_2011}.

\begin{figure}
	\includegraphics[height=3.2cm]{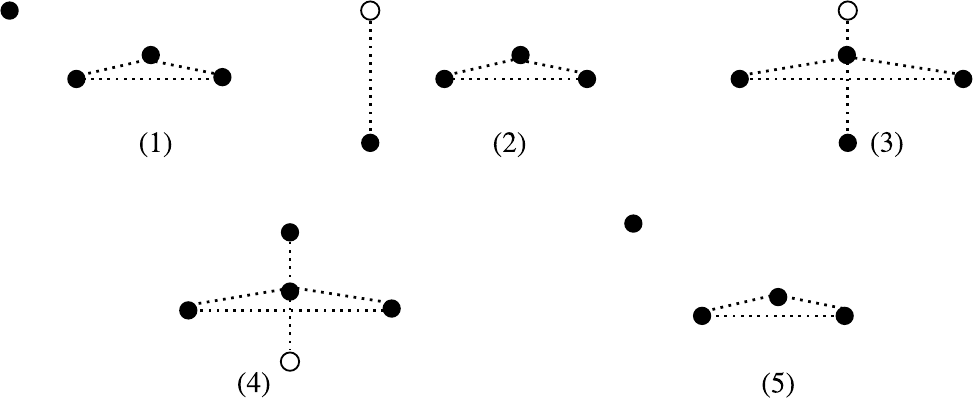}
	\caption{Sequence of steps leading to the notion of mutual statistics.} \label{mutual}
\end{figure}

We can define the notion of mutual statistics among a single excitation and a three-particle composite excitation \cite{Pai2019,Song:2023rml}. Let us denote such a state as $\ket{1,3}$, corresponding to the configuration in 1 of Fig. \ref{mutual}. First, we consider the transport of the single excitation along the $z$ direction through the application of a Wilson line operator without intercepting the triangular region defined by the three-particle excitation. Then, we consider the application of an open membrane operator to put apart the three excitations, enlarging the region defined by them. Carrying the reverse transport of the single excitation along the $z$ direction, now the line operator intersects the membrane and produces a minus sign. The final step is to apply the membrane operator again to return the three excitations to their initial position. This sequence of operations is depicted in Fig. \ref{mutual}. In algebraic terms, it reads
\begin{equation} \label{eq:mutual}
{M_I^{a}}^{\dagger} \, {W_I^{a}}^{\dagger} \, M_I^a \, W_I^a \ket{1,3} = - \ket{1,3}, 
\end{equation}
where we have used the algebra \eqref{projective}. The minus sign on the right hand side implies nontrivial mutual anyonic statistics among the involved excitations, which is a signature of long-range entanglement of topological ordered phases.

\section{Spontaneous Breaking of Subsystem Symmetries}

It is enlightening to view fractal topological order from the perspective of spontaneous breaking of subsystems symmetries. Topological order is characterized by the indistinguishability of the ground states, which means that for any {\it local operator} $\Phi$, it follows that
\begin{equation} \label{ind}
\bra{GS, a} \Phi \ket{GS,b}= C \delta_{ab},
\end{equation}
where $C$ is a constant independent of the particular ground state $\ket{GS,a}$. Distinct ground states are connected by {\it extended} symmetry operators. This is precisely the notion of spontaneous symmetry breaking, but for a generalized symmetry, where  {\it extended} symmetry operators act nontrivially on the ground states, taking from one to another. 

In order to ensure the property \eqref{ind}, we need two generalized symmetries mixed by a 't Hooft anomaly. Let us consider a specific pair of line $W$ and membrane $M$ operators, satisfying $[H,W]=[H,M]=0$ and $[M,W]\neq 0$, with $W^2=M^2=\openone$. We choose the ground states $\ket{GS,a}$, $a=1,2$, as simultaneous eigenstates of $H$ and $W$, 
\begin{equation} \label{eing}
H \ket{GS,a} = E_0 \ket{GS,a} ~~ \text{and} ~~ W \ket{GS,a} = \lambda_a \ket{GS,a},
\end{equation}
where $\lambda_a = \pm 1$ and $\lambda_2 = - \lambda_1$. Then, $\ket{GS,1}$ and $\ket{GS,2}$ are connected by the membrane operator, $\ket{GS,2} = M \ket{GS,1}$. 
Now let us see how the two symmetries $W$ and $M$ lead to \eqref{ind}. We start by computing  $\bra{GS,2} \Phi \ket{GS,2}= \bra{GS,1} M \Phi M \ket{GS,1}$.
As the membrane operator $M$ is mobile along the $z$ direction, we can move it in order to avoid the position of the local operator, which means that $M$ commutes with $\Phi$. This implies that  $\bra{GS,2} \Phi \ket{GS,2} = \bra{GS,1} \Phi \ket{GS,1}$.

\begin{figure}
	\includegraphics[height=3.2cm]{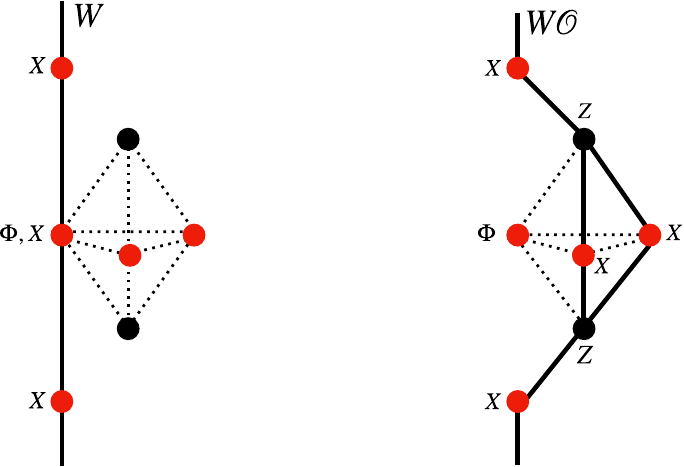}
	\caption{Local splitability of the Wilson line.} \label{split}
\end{figure}

It remains to show that $\bra{GS,2} \Phi \ket{GS,1} = 0$. We start with the left hand side of this equation and use \eqref{eing} to write it as $\bra{GS,2} \Phi \ket{GS,1} = \lambda_1^{-1} \lambda_2^{-1} \bra{GS,2} W \, \Phi \, W \ket{GS,1}$. We cannot use the same argument as before to justify the commutation of $W$ and $\Phi$ in the case where they intercept because $W$ is rigid. However, such a line operator enjoys a slightly different property, which we refer to as {\it local splitability}. This is a kind of nontopological deformation in the sense that it does not preserve the form of a line. Nervertheless, this can be used to avoid the point of support of the local operator. Supposing that $\Phi$ is located at a site $j$, we can use a TD operator $\mathcal{O}$ containing an $X_j$, to define the new object $\tilde{W} \equiv \mathcal{O} W = W \mathcal{O}$, which avoid the site $j$ at the price of splitting the line around it, as shown in Fig \ref{split}. With this, 
\begin{eqnarray}
	\bra{GS,2} \Phi \ket{GS,1} &=& \lambda_1^{-1}\lambda_2^{-1} \bra{GS,2} W \, \Phi \, W \ket{GS,1} \nonumber \\
	&=& - \bra{GS,2} \mathcal{O} \, W \, \Phi \, W \, \mathcal{O} \ket{GS,1} \nonumber \\
	&=&  - \bra{GS,2} \Phi \ket{GS,1},
\end{eqnarray} 
where we have used that $\mathcal{O}\ket{GR,a} = \ket{GR,a}$. Therefore, we obtain $\bra{GS,2} \Phi \ket{GS,1} = 0$. In conclusion, the two subsystem symmetries possessing a mixed 't Hooft anomaly ensure the condition of indistinguishability of the ground states required for topological order, which in turn is equivalent to the spontaneous breaking of the subsystem symmetries. 

\section{Explicit Breaking of Fractal Subsystem Symmetries}

For sizes $L$ as in Eq. \eqref{sizes} with $n = m + 1$ or sizes that cannot be expressed in the form of Eq. \eqref{sizes}, the ground state is unique since there are no nontrivial solutions for $t$'s. Accordingly, there are no membrane operators that commute with the Hamiltonian, since they do not close. In other words, the fractal subsystem symmetries are explicitly broken. Membrane operators that do not commute with a single TD operator can be used to create local excitations. This implies that there is only a single global anyonic superselection sector, as this type of membrane can be used to create and destroy single particles. The corresponding phases, however, are also topologically ordered as the notion of  mutual statistics in Eq. \eqref{eq:mutual} is still present.

While spontaneous symmetry breaking of discrete subsystem symmetry produces topological ordered phases with nontrivial ground state degeneracy, it is not a necessary condition (instead, it is a sufficient condition) as long as we define topological order in terms of long-range entanglement \cite{Wen_2017}. This is suitable for phases with a unique ground state, since the indistinguishability condition \eqref{ind} is trivial in such cases.  On the other hand, the mutual statistics in \eqref{eq:mutual} implies long-range entanglement and, consequently, topological order. 

Every Wilson line can be expressed in terms of a product of TD operators for these system sizes. Consider a membrane that creates a local excitation at the site $i$, $M_i^a$. According to the correspondence shown in Fig. \eqref{constraint_membrane}, we can construct from such a membrane an operator involving the product of $\mathcal{O}$'s, whose result is an isolated $X$ at the site $i$ and a set of $Z$'s in neighboring planes above and below $X$. The product of this structure along the $z$ direction annihilates the $Z$'s so that we end up with a Wilson line, $W_i^a = \prod_{m \, \in \, z}\prod_{p \, \in \, \mathcal{M}_i^a} \mathcal{O}_p^m$ \cite{Note1}. Acting on the ground state all these operators become the identity. These are trivial symmetry operators in the sense that they involve only operators that are present in the Hamiltonian.

If we cut open the system along the $xy$ plane at fixed coordinates $z = z_1$ and $z = z_{2}$, the TD operators at the cut planes will split in two and become boundary operators, $\mathcal{O} \rightarrow \mathcal{B}^{z_1} \, \mathcal{B}^{z_2}$. In this case, Wilson lines need to be attached to boundary operators,  
\begin{widetext}
\begin{equation} \label{wilson1}
	W_i^1 = \underbrace{ \left( \prod_{m \, \in \, z \neq z_0 }\prod_{ p \, \in \, \mathcal{M}_i^1}  \mathcal{O}_p^m \right) }_{\text{bulk}}
		\underbrace{ \left( \prod_{p \, \in \, \mathcal{M}_i^1} Z_p^{L - \frac{a_0}{2}} (X X X)_p^L \right) }_{ \text{boundary at } z_1}
		\underbrace{ \left( \prod_{p \, \in \, \mathcal{M}_i^1} 
		(X X X)_p^{z_0} Z_p^{z_0 - \frac{a_0}{2}} \right) }_{\text{boundary at }z_2}
\end{equation}  
and
\begin{equation} \label{wilson2}
	W_i^2 = \underbrace{ \left( \prod_{ m \, \in \, z - a_0/2 } \prod_{ p \, \in \, \mathcal{M}_i^2}  \mathcal{O}_p^m \right) }_{\text{bulk}}
	\underbrace{ \left( \prod_{p \, \in \, \mathcal{M}_i^2}  Z_p^{L} \right) }_{\text{boundary at }z_1}
		\underbrace{ \left( \prod_{p \, \in \, \mathcal{M}_i^2} Z_p^{z_0} \right) }_{\text{boundary at }z_2},
\end{equation}
\end{widetext}
for the lines in the two sublattices. These symmetry operators are no longer constrained to act as the identity on the ground state because the presence of the boundary operators. 

This leads to the existence of protected edge modes as long as the extended symmetries $W$ are preserved \footnote{This analysis is also valid for sizes of the form \eqref{sizes}. We leave for an upcoming publication a detailed analysis of boundary physics.}. While the Wilson lines \eqref{wilson1} and \eqref{wilson2} involving both bulk and boundary operators do commute among themselves, there is a nontrivial algebra among boundary operators. In other words, while the symmetries are realized exactly in the whole system (bulk+boundary), they are anomalous when considering only the boundary. Consider, for example, the boundary symmetry operators at $z = z_0$, $\mathcal{B}_i^1 \equiv \prod_{p \, \in \, \mathcal{M}_i^1} Z_p^{z_0 - \frac{a}{2}} (X X X)_p^{z_0}$ and $\mathcal{B}_i^2 \equiv Z_i^{z_0}$. They satisfy $\mathcal{B}_i^1 \, \mathcal{B}_j^2 = \left(-1\right)^{\delta_{ij} } \mathcal{B}_j^2 \, \mathcal{B}_i^1$, which leads to a nontrivial degeneracy of the boundary modes. Through the application of TD operators, these boundary operators can be stretched into the bulk, so that the nontrivial algebra flows to the bulk (anomaly inflow). 


\section{Conclusions}

We have reported on a model that presents exotic topological order due to an intricate interplay between fractal subsystem symmetries and the lattice size. 
The fractal subsystem symmetries are quite sensitive to the lattice size and exist only for linear sizes of the form $L = k \left( 2^n - 2^m \right)$, with integers $k \geq 1$, $ n > 2$, and $m \geq 0$, satisfying $n > m + 1$. The fractal symmetries possess mixed 't Hooft anomalies with the line subsystem symmetries, which lead to a nontrivial ground state degeneracy. This amounts to the spontaneous breaking of the fractal symmetry. For the remaining sizes, the fractal symmetry is explicitly broken and there are no fractal membrane operators that commute with the Hamiltonian. Consequently, there are no mixed 't Hooft anomalies and the ground state is unique. Despite the unique global anyonic superselection sector, such cases are topologically ordered, since the ground state is still long-range entangled.
		

\begin{acknowledgments}
	We are grateful to Huan Souza, who was involved in the initial stage of the work, Renann Jusinskas, Yizhi You, and Shiyu Zhou for enlightening discussions and comments on the draft. This work was partially completed during the Paths to Quantum Field Theory 2023 workshop at Durham University (G.D.). It is supported by DOE Grant No. DE-FG02-06ER46316 (G.D.) and Brazilian funding agencies CAPES and CNPq.
\end{acknowledgments}	
	
	
\bibliography{CCmodel_final_refs}

\setcounter{secnumdepth}{2}

\renewcommand{\appendixname}{Supplementary Material}
\appendix

\include{sup_material}

\end{document}

%% file: sup_material.tex
\title{Supplementary Material for \\
	``Fractal Subsystem Symmetries, `'t Hooft Anomalies, and UV/IR Mixing'' \label{SM}}
	
\begin{abstract}
	In this supplementary material, we provide additional insights and details pertaining to the discussions of ``Fractal Subsystem Symmetries, 't Hooft Anomalies,  and UV/IR Mixing.''  In Sec. \ref{MS1}, we calculate the ground state degeneracy and present illustrative examples. In Sec. \ref{MS2}, we demonstrate that the lowest value of $k$ yields the maximum degeneracy. In Sec. \ref{MS3}, we determine the number of independent Wilson line operators. In Sec. \ref{MS4} we discuss how to obtain the algebra among fractal membranes and Wilson line operators. Finally, in Sec. \ref{MS5}, we establish that Wilson lines can be simplified into a product of TD operators for non-degenerate system sizes.
\end{abstract}

\maketitle

\addcontentsline{toc}{section}{Supplementary Material}

\section{Ground State Degeneracy} \label{MS1}

The goal here is to find the ground state degeneracy as a function of the size of the lattice, which means to determine the number of constraints $N_C$. To systematically study the constraints of the model,  we follow the strategy proposed in \citealp{Bravyi_2011}. 

We start by writing the set of constraints generically as,
\begin{equation} \label{eq:set}
	\prod_{p \, \in \, \Lambda} \mathcal{O}_{p}^{t_{p}} = \openone,
\end{equation}
involving a set of weights $t_p$ mod 2 and with $\Lambda$ corresponding to the lattice defined by the points of the center of  TD operators. The number of nontrivial and linearly independent solutions for the set  $\{t_p\}$ gives the number of independent constraints $N_c$ of the model. 

\begin{figure}
	\centering
	\includegraphics[height=4cm]{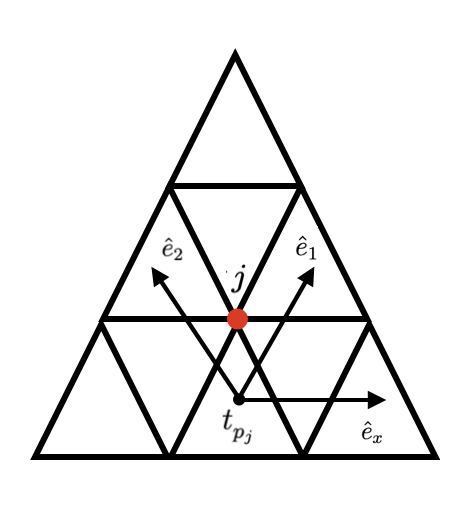}
	\caption{The $xy$-plane.} \label{new_basis}
\end{figure}

By rearranging the product in \eqref{eq:set} in terms of the product over the lattice sites where the spin operators are located $\tilde{\Lambda}$ (instead of the original $\Lambda$), we see that at each site there is the contribution of five TD operators, 
\begin{equation} \label{spin_lattice}
	\prod_{p \, \in \, \Lambda} \mathcal{O}_{p}^{t_{p}}
	= \prod_{j \, \in \, \tilde{\Lambda}} Z_j^{t_{p_j - {\hat{z}}/{2}}} X_j^{t_{p_j}}
	X_j^{t_{p_j + \hat{e}_1}}
	X_j^{t_{p_j + \hat{e}_2}}
	Z_j^{t_{p_j + {\hat{z}}/{2}}}
	= \openone,
\end{equation}
where for each site $j \in \tilde{\Lambda}$ we have a corresponding operator $\mathcal{O}_{p_j}$ at $p_j\in \Lambda$ as a reference, and $\hat{e}_1$ and $\hat{e}_2$ are vectors in the $xy$-plane locating the neighboring operators, as shown in Fig. \ref{new_basis}.

Since the product of different Pauli matrices is not the identity, the only way to satisfy the equation (\ref{spin_lattice}) is
\begin{equation} \label{equality1}
	t_{p_j + \frac{\hat{z}}{2}} + t_{p_j - \frac{\hat{z}}{2}} = 0 \pmod{2}
\end{equation}
and, simultaneously,
\begin{equation} \label{equality2}
	t_{p_j} + t_{p_j + \hat{e}_1} + t_{p_j + \hat{e}_2} = 0 \pmod{2}.
\end{equation}
Eq.~\eqref{equality1} comes from the $Z$-operators and implies that all TD operators in the same line along the $z$-direction are subjected to the same weight. 
So, the whole set $\{t_p\}$ can be uniquely defined by two planes depicted in the Fig. 1 of the main text, namely, one plane of the sub-lattice $\Lambda_1$, containing red triangles, and the other of the sub-lattice $\Lambda_2$, containing black ones.
On the other hand, Eq.~\eqref{equality2} describes the constraints in a $xy$-plane, which are related to the underlying fractal symmetry. 

\begin{figure}
	\centering
	\includegraphics[width=8cm]{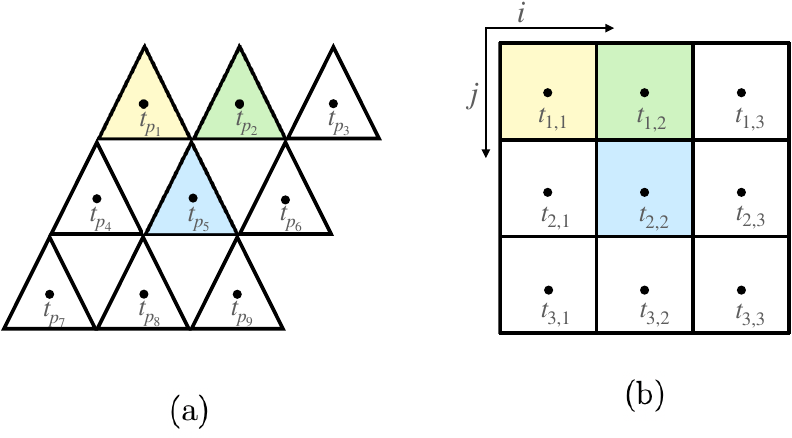}
	\caption{In Figure (a) the positions of the weights are represented within the triangular lattice, while Figure (b) shows the redefined positions, ensuring compliance with periodic boundary conditions.} \label{new_position}
\end{figure}

It is convenient to express \eqref{equality2} in a slightly different way by associating a matrix element to each one of the $t$'s partaking the constraint, so that we rewrite \eqref{equality2} as
\begin{equation} \label{eq:coefficients}
	t_{j + 1,i} = t_{j,i} + t_{j,i - 1} \pmod{2}.
\end{equation}
This is illustrated in an example in Fig. (\ref{new_position}), with the periodic boundary condition $t_{j, i + L} = t_{j, i}$. 

Any configuration of $t$'s along a line completely determines the subsequent lines in the $xy$-plane. To understand this, it is convenient to represent a line configuration in terms of a polynomial \cite{,Yoshida:2013sqa,Devakul_2019}
\begin{equation} \label{eq:Laurent}
	S_{j}(x) = \sum_{i = 1}^{L} t_{j,i} x^{i},
\end{equation} 
whose coefficients are defined mod 2. Accordingly, the periodic boundary condition $t_{j, i + L} = t_{j, i}$ and $t_{j + L, i}=t_{j,i}$ translates into $x^{i + L} = x^i$ and $S_{j + L} (x) = S_{j} (x)$, respectively. The polynomial representation \eqref{eq:Laurent} does not embody automatically the constraint \eqref{line}, but we shall see how to implement it in a moment.  

Now, the local constraint in  \eqref{eq:coefficients} allows us to determine the update rule for the subsequent line according to
\begin{equation}
	S_{j + 1} (x) = \sum_{i = 1}^{L} \left(t_{j,i} + t_{j, i - 1} \right) x^{i} = \left( 1 + x \right) S_j(x).
\end{equation}
Applying it recursively, we find
\begin{equation} \label{generic_line}	
	S_{j + 1} (x) = \left( 1 + x \right)^{j }S_{1} (x).
\end{equation}
Therefore, given the configuration of the first line $S_1 (x)$, this relation enables us to determine the configuration of the whole $xy$-plane. 

\begin{figure}
	\includegraphics[width=0.36\linewidth]{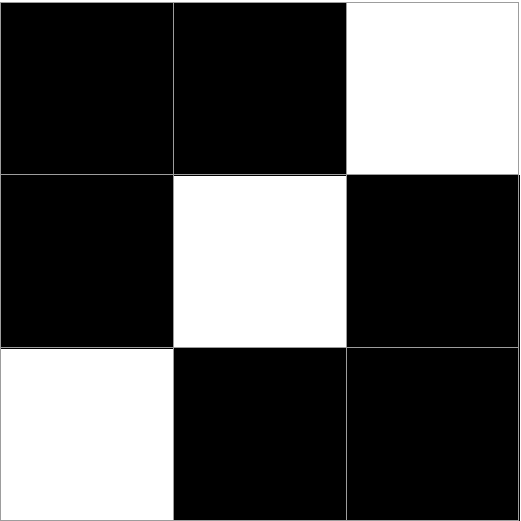} \hspace{0.1\linewidth}
	\includegraphics[width=0.36\linewidth]{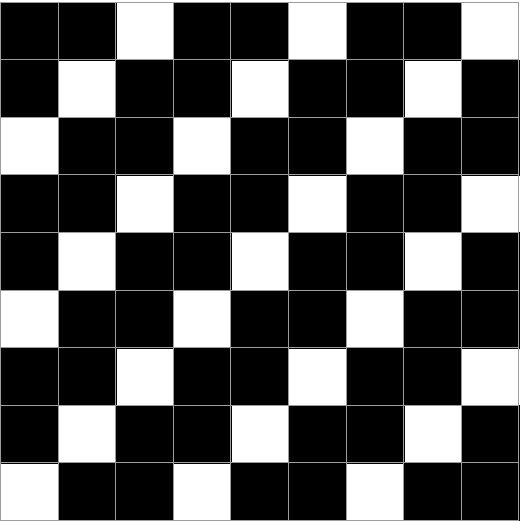} \\ \vspace{0.1\linewidth}
	\includegraphics[width=0.36\linewidth]{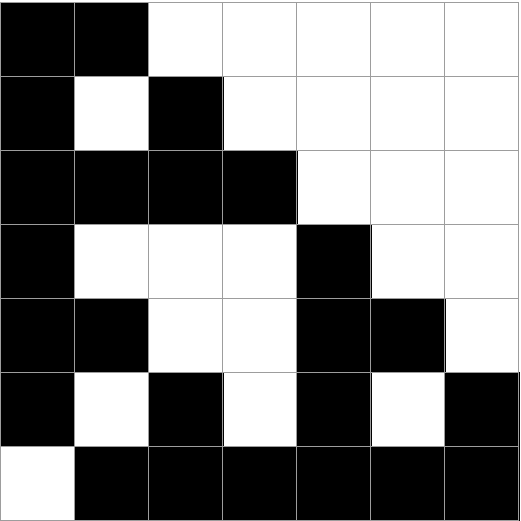} \hspace{0.1\linewidth}
	\includegraphics[width=0.36\linewidth]{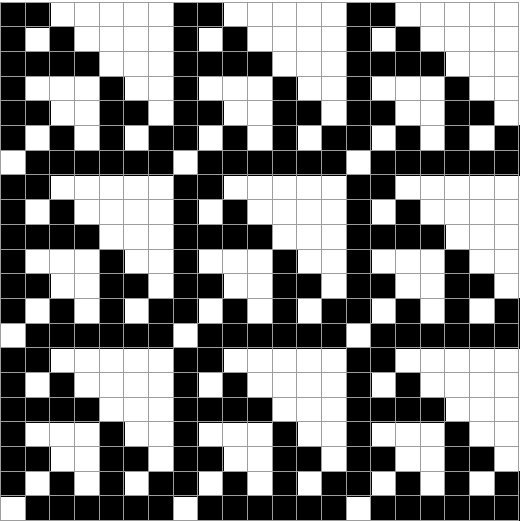} 
	\caption{The constraints for system sizes $L = 3$, $L = 9$, $L = 7$, and $L = 21$ respectively, in which the black squares are equivalent to $t_{ij} = 1$.} 	\label{fig:constraints}
\end{figure}

We can iterate the local constraint \eqref{eq:coefficients} to generate an extended constraint along a line
\begin{equation} \label{line}
	t_{j,1} + t_{j,2} + t_{j,3} + \cdots + t_{j, L} = 0 \pmod{2},
\end{equation}
which implies that a generic line must have an even number of elements with $t_{j,i} = 1$ to satisfy the constraint. 
A basis for configurations satisfying \eqref{line} is constituted of a set of  polynomials of the form $x^p + x^q$, for integers $p$, $q \geq 0$, and $p \neq q$. The question is how to determine the number of independent elements in this basis for a given lattice size. This counting can be systematized by introducing a factor of $( 1 + x )^{2^m}$ into the definition of the polynomial $S_1 (x)$,
\begin{equation} \label{first_line}
	S_{1} (x) = \left( 1 + x \right)^{2^{m}}\sum_{i = 1}^{L} t_{1, i} x^{i},
\end{equation}
where $m = 0, 1, 2, \, \ldots \,$. For any choice of $t$'s and $m$ the expression \eqref{first_line} generates either an element of the basis or a linear combination of them, and therefore it satisfies the constraint \eqref{line}. This is so because $( 1 + x )^{2^m} = 1 + x^{2^m}$, which follows from the fact that all coefficients of $S_{j} (x)$ are evaluated mod 2, and all coefficients $\binom{2^m}{l}$ in the binomial expansion are divisible by two (i.e., $\binom{2^m}{l} \bmod{2} = 0$) except those with $l = 0$ and $l = 2^m$.  At the same time, the relation \eqref{first_line} is convenient since it provides a simple update rule according to \eqref{generic_line}, namely, 
\begin{equation} \label{j+1_line}
	S_{j + 1} (x) = \left( 1 + x \right)^{j + 2^{m}}\sum_{i = 1}^{L} t_{1,i} x^{i}.
\end{equation}
The requirement that this expression satisfy the periodic boundary condition $S_{L + 1} (x) = S_1 (x)$ implies that
\begin{equation}
	L+ 2^m = 2^n,
\end{equation}
where $n = 2, 3, \, \ldots \,$. In other words, only those linear sizes $L$ that can be expressed in the form
\begin{equation} \label{sizes1}
	L = 2^n - 2^m,
\end{equation}
with $n > m \geq 0$, are compatible with periodic boundary conditions and admit nontrivial solutions for $t$'s, leading to a nontrivial ground state degeneracy. Two examples of the Sierpinski fractal structure following from \eqref{first_line} is shown in Fig. \ref{fig:constraints} for $L = 3$ and $L = 7$.

For the lattice sizes given in \eqref{sizes1}, we can determine the number of independent elements in the basis for the polynomials satisfying \eqref{line}. To see this, we choose a configuration  of $t$'s in \eqref{first_line} so that a single $t$ is nonvanishing, 
\begin{equation} \label{001}
	S_1^i (x) = (1 + x^{2^m}) x^i.
\end{equation}
Because of the periodic boundary condition $x^{i + L} = x^i$, we find that for 
\begin{equation}
	i = 1, 2, \, \ldots \, , 2^n - 2^{m + 1},
\end{equation}
the relation \eqref{001} produces linearly independent elements. In other words, this implies a total number of $2^n - 2^{m + 1}$ constraints for each sublattice. Taking into account that there are two independent sublattices, we obtain finally 
\begin{equation}
	N_c = 2 (2^n - 2^{m + 1}),
\end{equation}
and consequently the ground state degeneracy 
\begin{equation} \label{gsd1MS}
	GSD= 2^{(2^{n + 1} - 2^{m + 2})}.
\end{equation}
It follows that for $n > m + 1$, the ground state is not unique and the system is topological ordered. 


This degeneracy is associated with the fractal symmetry of the model and is comparable to that of other models sharing the same symmetry. The results we had obtained are aligned with those reported for the Triangular Plaquette Model in \cite{Sfairopoulos2023}. Both results can be compared, taking into account the degeneracy resulting from the number of constraints in a single plane, particularly for cases where $L_x = L_y$.


\section{Case $L = k( 2^n - 2^m ) = k' ( 2^{n'} - 2^{m'} )$} \label{MS2}

We discuss here an issue that arises for sizes $L = k ( 2^n - 2^m )$. It may happen that there is more than one way to express a given size in terms of copies, i.e., $L = k (2^n - 2^m) =  k' (2^{n'} - 2^{m'})$. In this case, we shall look for the buiding-block copy that provides the maximum number of nontrivial solutions for $t$'s, namely, which number among $2^n - 2^{m + 1}$ and $2^{n'} - 2^{m' + 1}$ is greater. 


To understand this for a general case,  we first note that the number $2^n - 2^{m + 1}$ in terms of $n$ and $k$ is expressed as
\begin{equation}
	2^n - 2^{m + 1} = \frac{2 L}{k} - 2^{n}. 
\end{equation}
Then, consider $n' = n - \epsilon$ and $k' = \lambda k$, where  $\epsilon = 1, 2, \,\dots \, , n - 2$ and $\lambda$ is a positive number,  so that $L = k \left( 2^{n} - 2^{m} \right) = \lambda k \left( 2^{n - \epsilon} - 2^{m'} \right)$. In order to satisfy $2^n - 2^{m + 1} > 2^{n'} - 2^{m' + 1}$, we find
\begin{equation}
	\frac{2 L}{k} - 2^{n} > \frac{2 L}{\lambda k} - 2^{n - \epsilon},
\end{equation}
which can be expressed equivalently as
\begin{equation}
	\lambda > \frac{1}{1 - \frac{2^{n - 1} k}{L} \left(1-2^{- \epsilon} \right)}.
\end{equation}
Since the right-hand side is greater than $1$ (since $\frac{2^{n - 1} k}{L} < 1$ and $1 - 2^{- \epsilon} < 1$), it follows that $ \lambda > 1$. Then, we find that $k < k'$ implies $2^n - 2^{m + 1} > 2^{n'} - 2^{m' + 1}$. To summarize, the maximum value of $2^{n} - 2^{m + 1}$ occurs for the smallest value of $k$.


\section{Independent Wilson Lines} \label{MS3}

\begin{figure}
	\includegraphics[height=5cm]{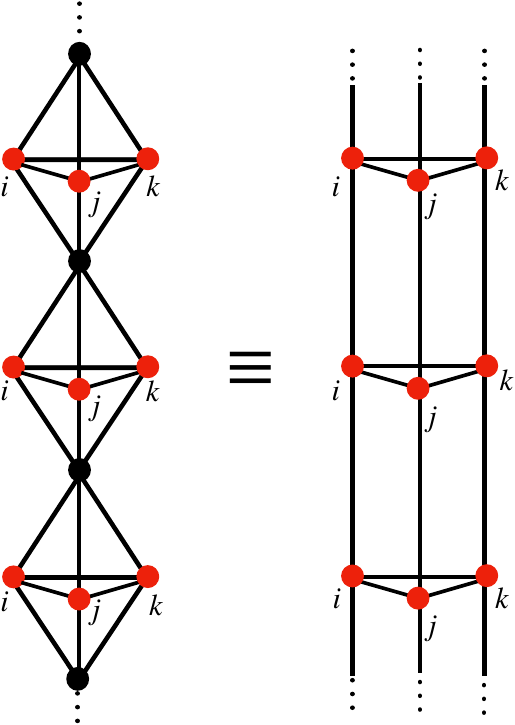}
	\caption{TD operators arranged along a closed line in the $z$-direction.} \label{tube}
\end{figure}

We want to determine the number of independent Wilson lines for sizes $L = k(2^n - 2^m)$. Consider the product of all operators $\mathcal{O}$ belonging to the sublattice $a$, acting on the ground state
\begin{equation} \label{lec}
	\prod_{p \, \in \, \Lambda^a} \mathcal{O}_p \ket{GS} = \ket{GS}. 
\end{equation}
We can think of this as a low-energy constraint. Now let us rewrite this expression in a more useful form, by separating from the product a set of TD operators arranged along a closed line in the $z$-direction, as in Fig. \ref{tube}. Writing this product of operators explicitly, we see that 
\begin{equation} \label{tubeop}
	\begin{aligned}
		\prod_{m \, \in \, z^a} \mathcal{O}_{ijk}^m &= \prod_{m \, \in \, z^a} Z^{m + a/2} \, X_i^m \, X_j^m \, X_k^m \, Z^{m - a/2}  \\
		&=  \left( \prod_{m \, \in \, z^a} X_i^m \right) \left( \prod_{m \, \in \, z^a} X_j^m \right) \left( \prod_{m \, \in \, z^a} X_k^m \right)  \\
		&= W_i^a \, W_j^a \, W_k^a.
	\end{aligned}
\end{equation}

Therefore, such a product of TD operators is equivalent to the product of three Wilson lines. With this, the constraint \eqref{lec} can be written as
\begin{widetext}
	\begin{equation}
		W_i^a \, W_j^a \, W_k^a   \sideset{}{'}\prod_{p \, \in \, \Lambda^a} \mathcal{O}_p \ket{GS} = \ket{GS} \qquad \Rightarrow \qquad W_i^a \, W_j^a \, W_k^a \ket{GS} = \ket{GS},
	\end{equation}
\end{widetext}
where the prime in the product means that is does not include the operators that were separated. The relation in the right shows that the three Wilson lines $W_i^a$, $W_j^a$, and $W_k^a$ are not independent.

Now, as the position of the operators in \eqref{tubeop} is arbitrary in the $xy$-plane, each possible position will lead to a new constraint among three lines. Thus, we end up with as many constraints as the number of operators in the plane, which in turn is the same as the number of sites of one sublattice, namely, $L \times L$. However, these $L^2$ constraints are not all linearly independent because of the topological constraints (see Eq. (3) of main text). Let us consider one of the constraints and write it as $\prod_{m \, \in \, z^a} \prod_{p \, \in \, \mathcal{M}_I^a} \mathcal{O}_p^m = \openone$. Then, we can separate the specific $p$ corresponding to the operator $\mathcal{O}_{ijk}^m$, to write this expression as
\begin{equation}
	\left( \prod_{m \, \in \, z^a}\mathcal{O}_{ijk}^m\right)  \prod_{m \, \in \, z^a}  \prod_{_{p \, \neq \, ijk}^{p \,\in \, \mathcal{M}_I^a}} \mathcal{O}_p^m = \openone.
\end{equation}
Using again relation \eqref{tubeop}, this implies that 
\begin{equation}
	W_i^a \, W_j^a \, W_k^a = \prod_{m \, \in \, z^a}  \prod_{_{p \, \neq \, ijk}^{p \, \in \, \mathcal{M}_I^a}} \mathcal{O}_p^m,
\end{equation}
namely, the constraint over $W_i^a$, $W_j^a$, and $W_k^a$ can be expressed in terms of other constraints among Wilson lines given by the right hand side of the above equation. We can repeat this process for each one of the $2^n - 2^{m + 1}$ constraints, so that the total number of independent constrains is $L^2 - (2^n - 2^{m + 1})$. Finally, the total number of independent Wilson lines is 
$L^2 - [ L^2 - (2^n - 2^{m + 1})] = 2^n - 2^{m + 1}$, which is precisely the number of membrane operators. Thus, we label the Wilson lines with the same type of index as the membrane operators,
\begin{equation} \label{WL1a}
	W_{I}^a = \prod_{j \, \in \, \mathcal{L}_{I}^a} X_j,
\end{equation}
where $\mathcal{L}_{I}^a$, with $I = 1, \, \ldots\, , 2^n - 2^{m + 1}$, corresponds to each one of the independent Wilson lines.  


\section{Pairwise Decoupling of Operators} \label{MS4}

We discuss here how to obtain the algebra among fractal membranes and Wilson line operators (see Eq. (9) of main text). It is enough to consider the first lines of the membrane operators associated with the configuration generated by \eqref{001}, which we rewrite here for convenience,
\begin{equation}
	S_1^i (x) = (1 + x^{2^m}) x^i, \qquad i = 1, \, \ldots \, , 2^n - 2^{m + 1}.
\end{equation}
Then we consider the set of $2^n - 2^{m + 1}$ linear combinations
\begin{equation} \label{lc}
	\tilde{S}_p^l (x) \equiv \sum_{i = 1}^l S_1^{ 2^m i - p},
\end{equation}
in which $p = 0, 1, \, \ldots \, , 2^m - 1$ and $l = 1, 2, \, \ldots\, , (2^n - 2^{m + 1})/2^m$. The associated membrane operators generated from these linear combinations have the following important property: each membrane operator contains one site occupied by a $Z$-operator that is not occupied by any other membrane operator. In this way, if we consider a Wilson line intercepting this point, it will anti-commute with the corresponding membrane operator and will commute with all the remaining ones. 

\begin{figure*}
	\includegraphics[width=0.12\linewidth]{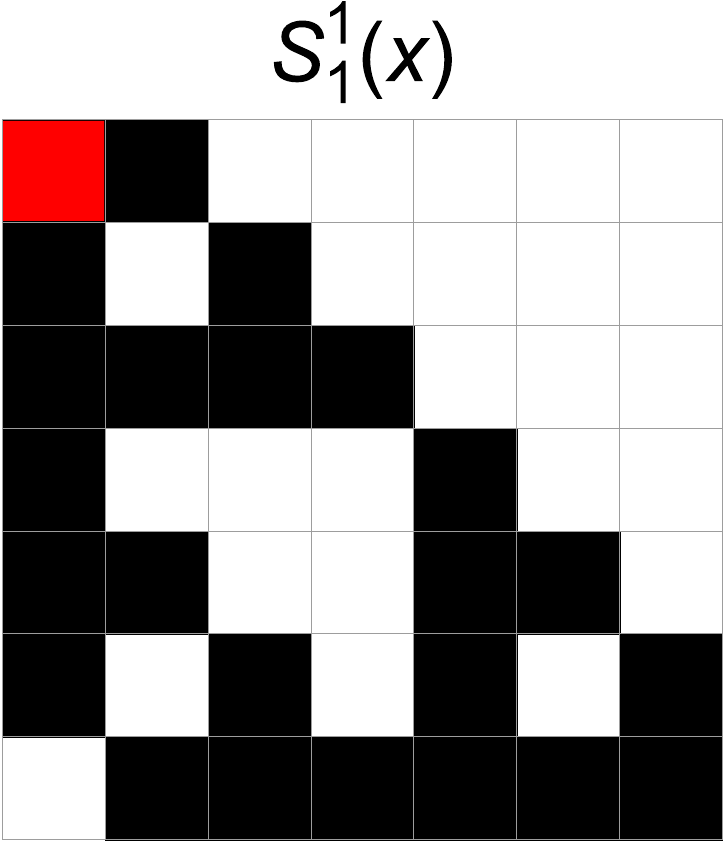} \hspace{0.01\linewidth} \includegraphics[width=0.12\linewidth]{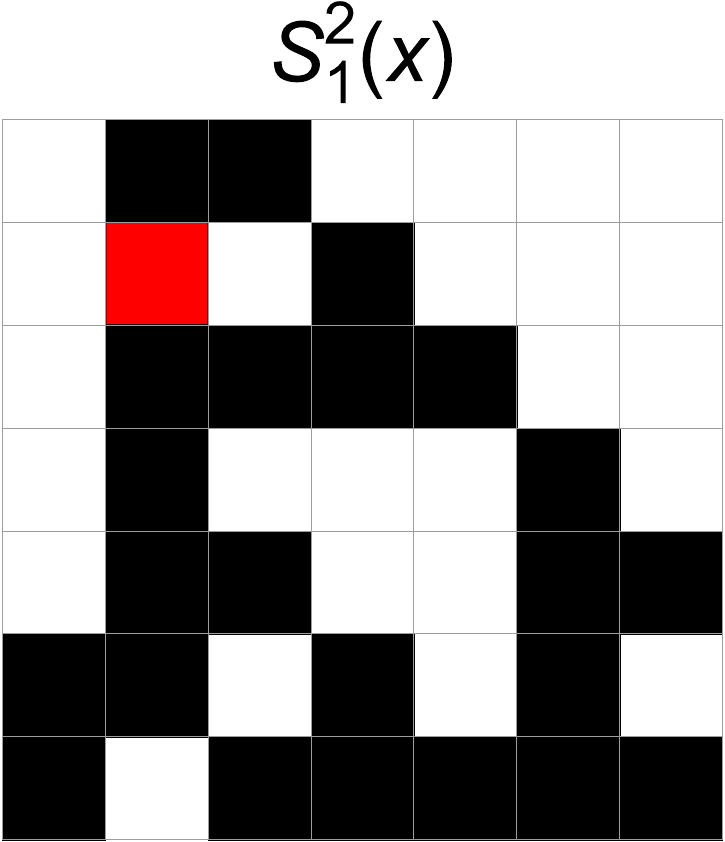} \hspace{0.01\linewidth} \includegraphics[width=0.12\linewidth]{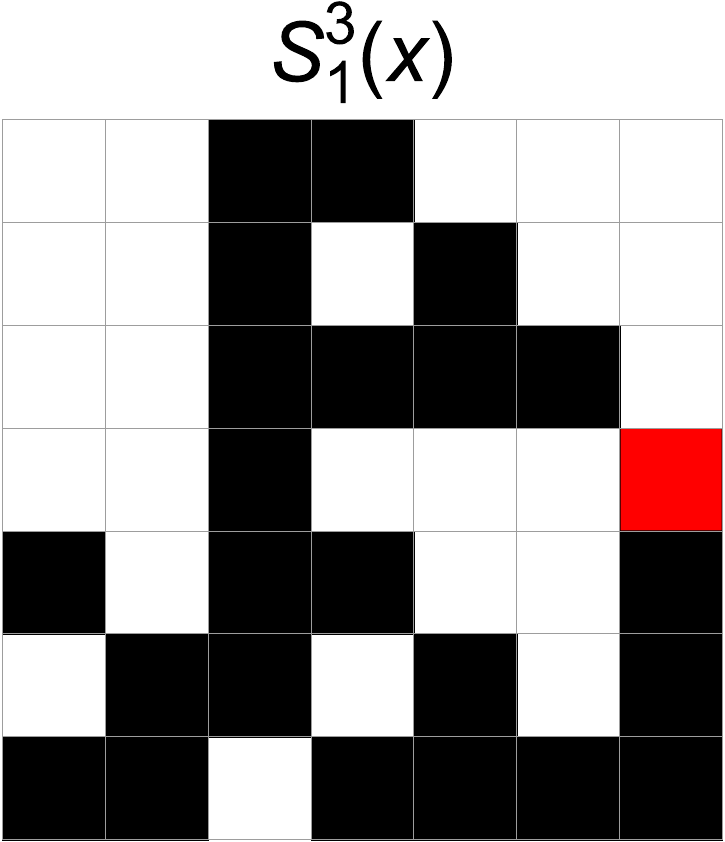} \hspace{0.01\linewidth} \includegraphics[width=0.12\linewidth]{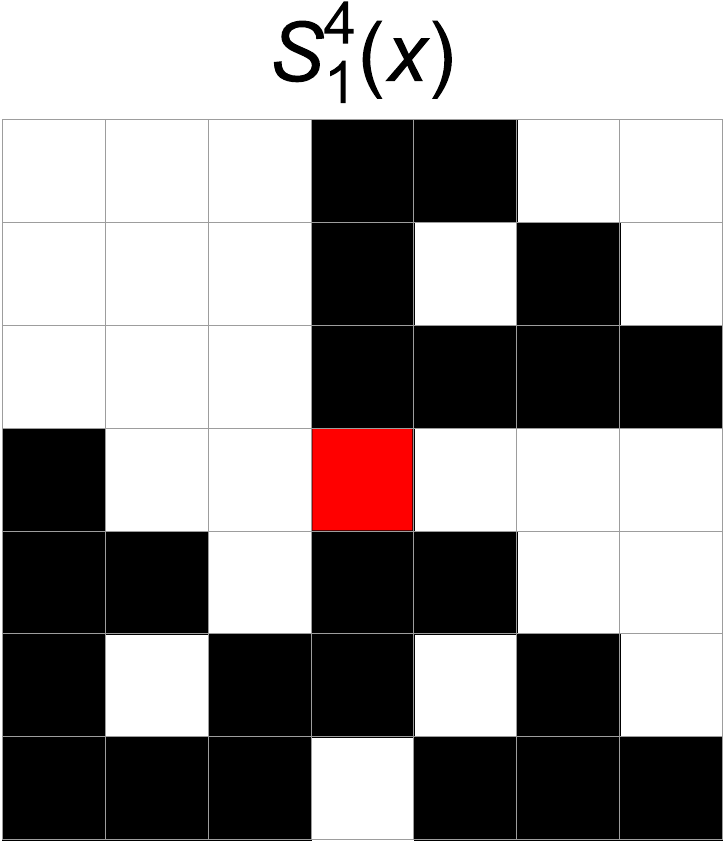} \hspace{0.01\linewidth} \includegraphics[width=0.12\linewidth]{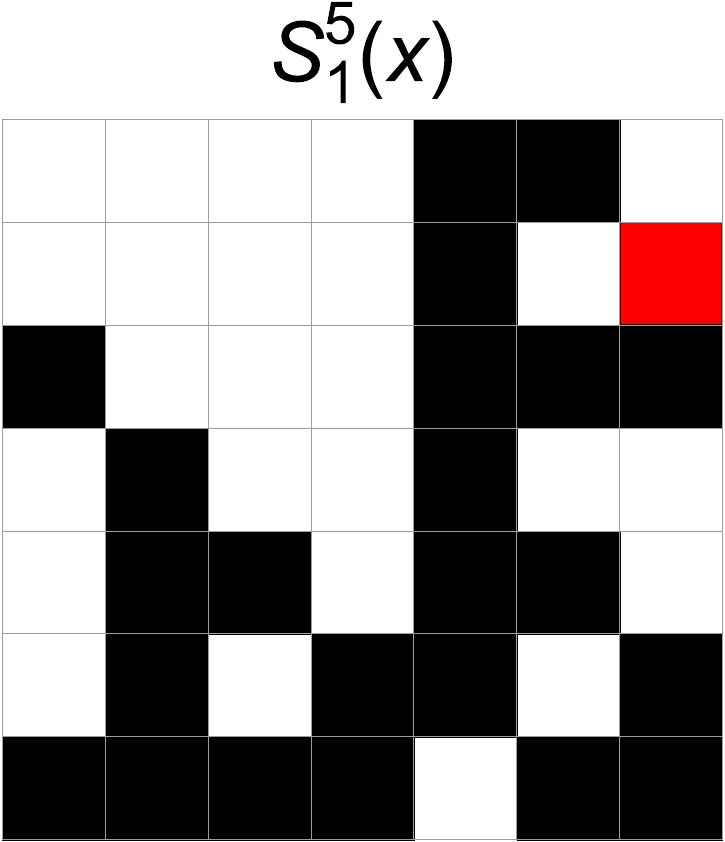} \hspace{0.01\linewidth} \includegraphics[width=0.12\linewidth]{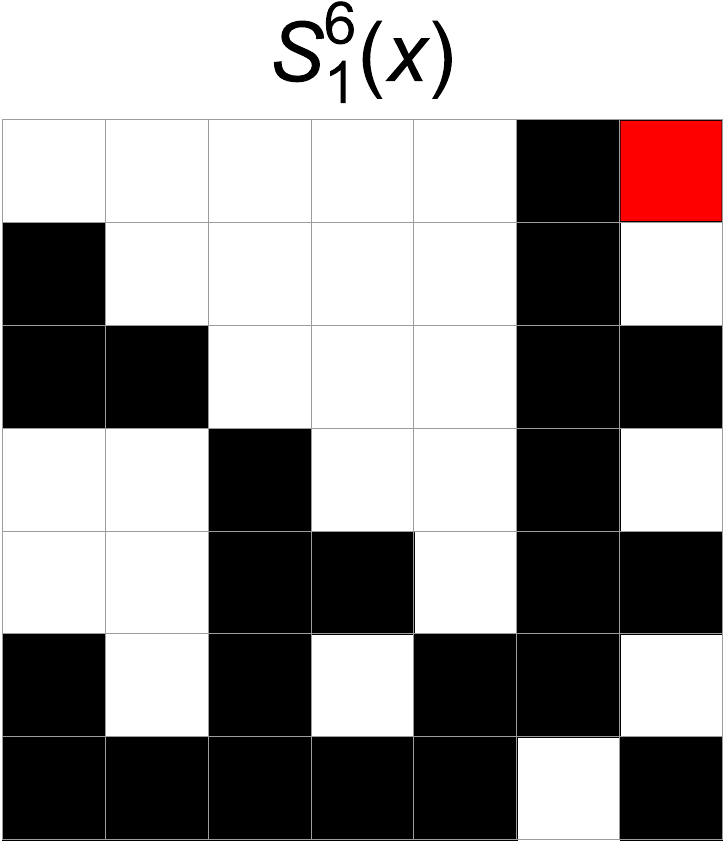} \\ \vspace{15pt}
	\includegraphics[width=0.12\linewidth]{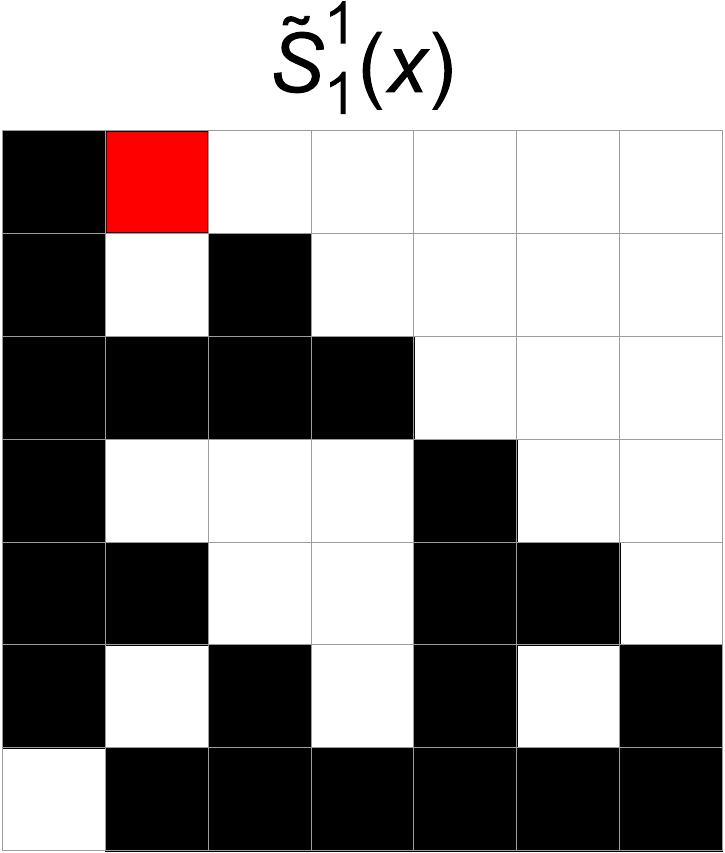} \hspace{0.01\linewidth} \includegraphics[width=0.12\linewidth]{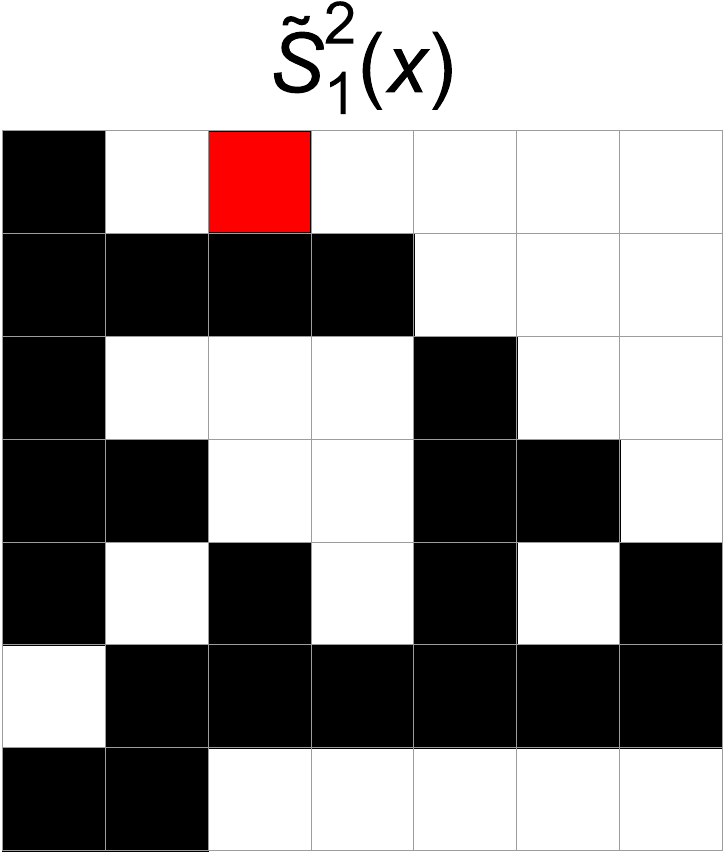} \hspace{0.01\linewidth} \includegraphics[width=0.12\linewidth]{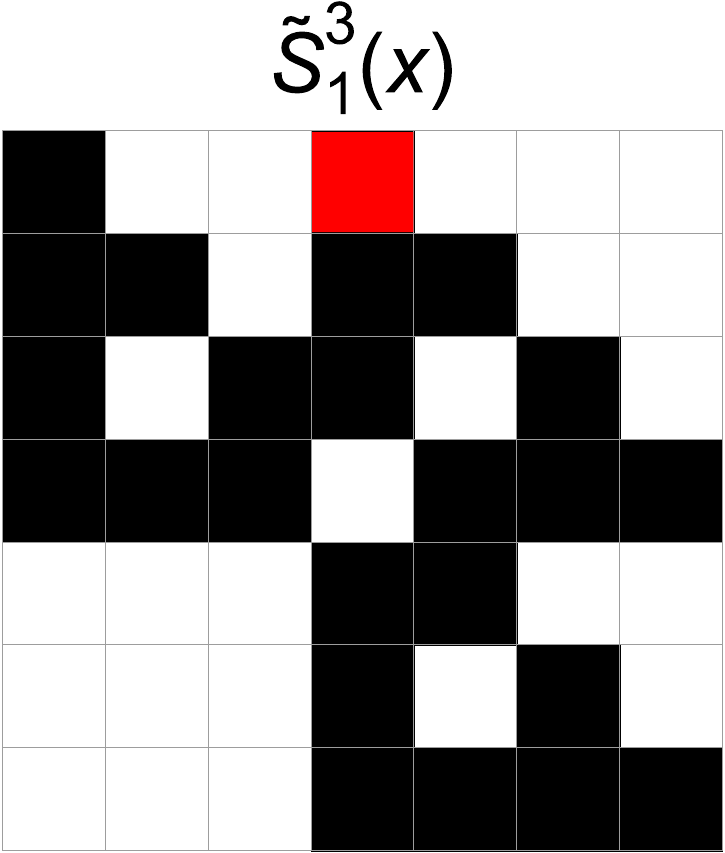} \hspace{0.01\linewidth} \includegraphics[width=0.12\linewidth]{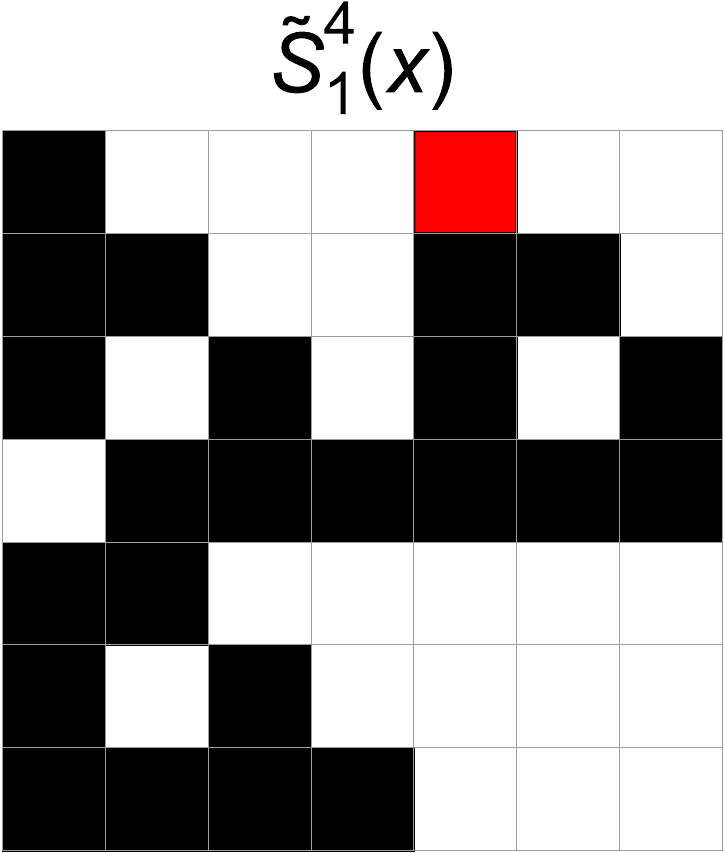} \hspace{0.01\linewidth} \includegraphics[width=0.12\linewidth]{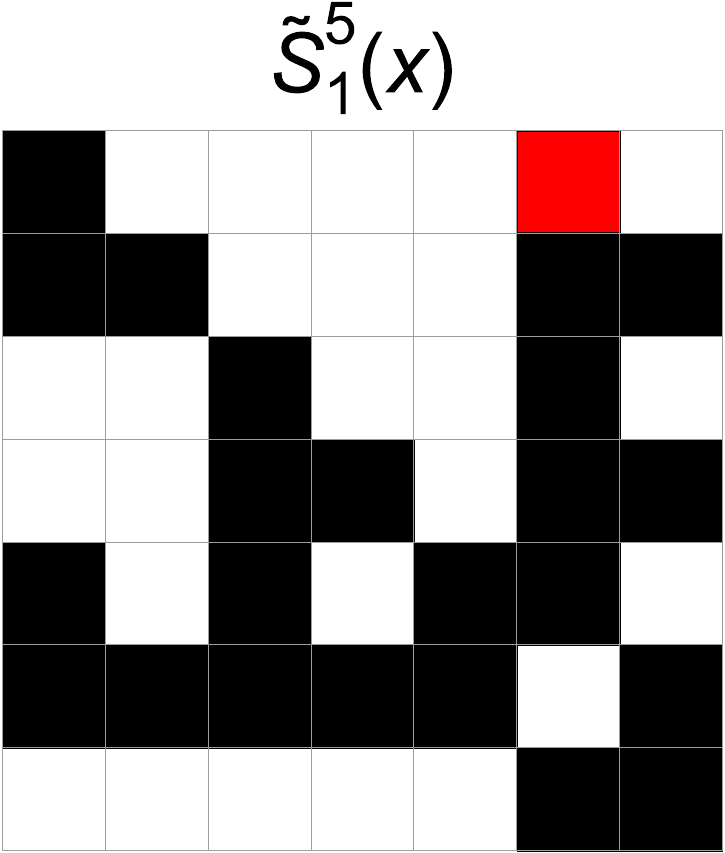} \hspace{0.01\linewidth} \includegraphics[width=0.12\linewidth]{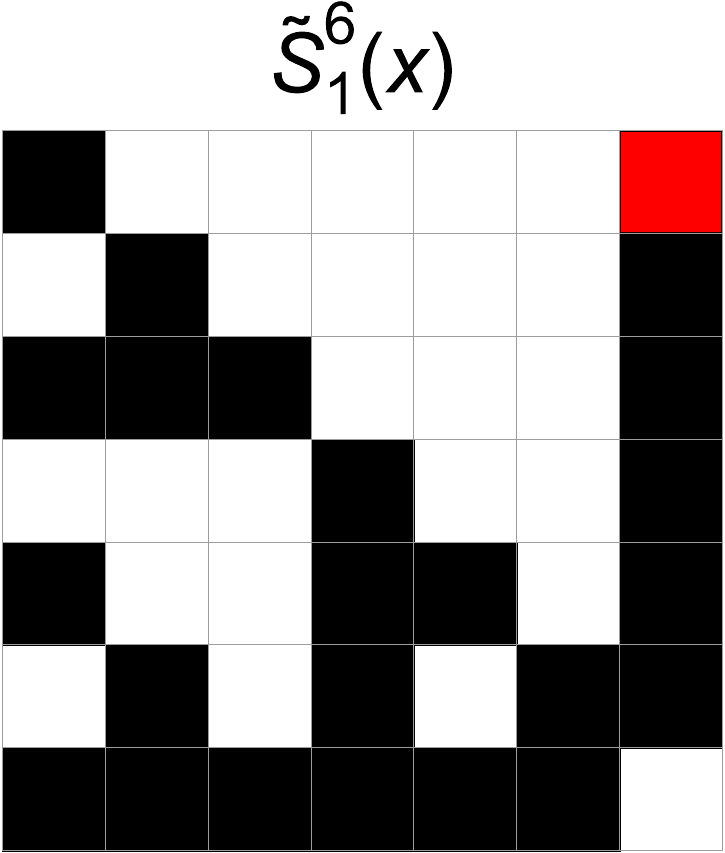}
	\caption{The membranes generated by $S_{1}^{i}(x)$ and by $\tilde{S}_{1}^{i}(x)$, with $i=1,2, \, \dots \, ,6$, in system size $L=7$. In each membrane it is highlighted the $t_{ij} = 1$ that do not repeat for the other membranes.} \label{fig:tildemembranes}
\end{figure*}

To illustrate this, we consider a specific example, say $n = 3$ and $m = 0$,  which corresponds to the size $L = 7$. We have the following set of independent configurations
\begin{equation}
	\begin{aligned}
		S_1^1 (x) &= x + x^2 \\
		S_1^2 (x) &= x^2 + x^3 \\
		S_1^3 (x) &= x^3 + x^4 \\
		S_1^4 (x) &= x^4 + x^5 \\
		S_1^5 (x) &= x^5 + x^6 \\
		S_1^6 (x) &= x^6 + x^7.
	\end{aligned}
\end{equation}
We see that with the exception of the sites $1$ and $7$, all the other sites in the first line configurations are occupied by $Z$-operators of two membranes (see Fig. \ref{fig:tildemembranes}).

For example, the site $3$ is occupied by the membrane operators constructed from the configurations $S_1^2 (x)$ and $S_1^3 (x)$. Thus, a Wilson line intercepting this site will anti-commute with both membrane operators, and commute with the remaining ones. Now, let us construct the linear combinations according to \eqref{lc}. We have
\begin{widetext}
	\begin{equation}
		\begin{aligned}
			\tilde{S}_0^1 (x) &= S_1^1 (x) = x + x^2 \\
			\tilde{S}_0^2 (x) &= S_1^1 (x) + S_1^2 (x) = x + x^3 \\
			\tilde{S}_0^3 (x) &= S_1^1(x) + S_1^2 (x) + S_1^3 (x) = x + x^4 \\
			\tilde{S}_0^4 (x) &= S_1^1 (x) + S_1^2 (x) + S_1^3 (x) + S_1^4 (x) = x + x^5 \\
			\tilde{S}_0^5 (x) &= S_1^1(x) + S_1^2 (x) + S_1^3 (x) + S_1^4 (x) + S_1^5 (x) = x + x^6 \\
			\tilde{S}_0^6 (x) &= S_1^1 (x) + S_1^2 (x) + S_1^3 (x) + S_1^4 (x) + S_1^5 (x) + S_1^6 (x) = x + x^7.
		\end{aligned}
	\end{equation}
\end{widetext} 
We see that each one of the sites $2,\, \ldots \, , 6$ is occupied only once by the $Z$-operators of the membrane operators associated to the above configurations. Then, by considering six Wilson line operators intercepting these six sites, we obtain the algebra in the form
\begin{equation} \label{projective}
	M_I^a \, W_J^b = \left(-1\right)^{\delta_{IJ} \delta_{ab}} W_J^b \, M_I^a. 
\end{equation}


\section{Wilson Lines as a Product of TD Operators} \label{MS5}

For sizes that explicitly break the fractal subsystem symmetry, every Wilson line operator can be expressed in terms of a product of TD operators. We provide here details of this construction. 

The first step to construct a Wilson line is to find a product of TD operators lying in a $xy$-plane that produces a single $X$ somewhere in the plane. At the same time, such a product introduces many $Z$-operators above and below the chosen plane. Stacking this product along the $z$-direction will eliminate the $Z$-operators and will produce a Wilson line. 

Let us consider a product of TD operators constructed in the following way. We take a first line configuration $S_1 (x) = \sum_{i = 1}^L t_{1, i} x^i$, with the subsequent lines constructed according to the update rule \eqref{generic_line}. The corresponding product of TD operators results in identities between the lines, remaining $X$-operators only among $S_{1} (x)$ and $S_{L} (x)$.  The task is to find a configuration for the first line $S_{1} (x)$ satisfying
\begin{eqnarray} \label{eq:nonhomo}
	S_{1} (x) + S_{L + 1} (x)  = x^k,
\end{eqnarray}
for some site $k$, which implies an isolated $X$-operator in the plane.

\begin{figure}
	\includegraphics[width=\linewidth]{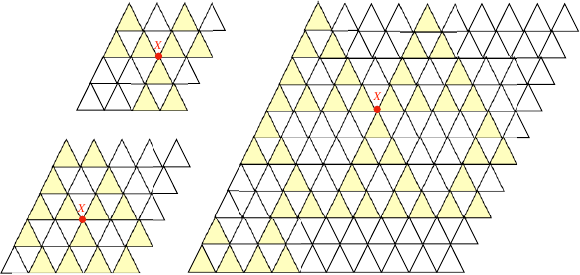}
	\caption{Membrane operators for sizes $L = 4$, $5$ and $10$ that lead to an isolated $X$-operator (considering periodic boundary conditions). The configurations of $S_{1} (x)$ are the solution of Eq. \eqref{eq:system} for each size, namely, $a_{1} = 1$ for $L = 4$, $a_{1} = a_{3} = a_{4} = 1$ for $L = 5$, and $a_{1} = a_{5} = a_{7} = 1$ for $L = 10$.}  \label{membraneoperator}
\end{figure}

The left hand side of \eqref{eq:nonhomo} can be written as
\begin{equation}
	\begin{aligned}
		S_{1} (x) + S_{L + 1} (x)  &= \left[ 1 + \left( 1 + x \right)^{L} \right] S_{1} (x) \\
		&= \sum_{i = 1}^{L}\sum_{j = 1}^{L} t_{1,i} \binom{L}{j} x^{i + j},
	\end{aligned}
\end{equation}
where not only $t_{1,i}$ but also the binomial coefficients are defined mod $2$. It is useful to redefine the sum in terms of $j' = i + j$ and use the identification $x^{j'} = x^{j' - L}$ for $j' > L$,
\begin{widetext}
	\begin{equation}
		\begin{aligned}
			\sum_{i = 1}^{L} \sum_{j' = i + 1}^{i + L} t_{1, i} \binom{L}{j' - i} x^{j'} &= \sum_{i = 1}^{L}t_{1, i} \left[ \sum_{j' = i + 1}^{L} \binom{L}{j' - i} x^{j'} + \sum_{j' = L + 1}^{i + L} \binom{L}{j' - i} x^{j' - L} \right]  \\
			&=  \sum_{i = 1}^{L} t_{1,i} \left[ \sum_{j = i + 1}^{L} \binom{L}{j - i} x^{j} + \sum_{j = 1}^{i} \binom{L}{j + L - i} x^{j} \right] \\
			&= \sum_{i = 1}^{L}\sum_{j = 1}^{L} t_{1,i}\binom{L}{\left\lvert i - j\right\rvert} x^{j}, 
		\end{aligned}
	\end{equation}
\end{widetext}
where in the last step we have used that $\binom{L}{L - (i - j)} = \binom{L}{i - j}$. With this, we can express \eqref{eq:nonhomo} as
\begin{equation} \label{eq:intersction}
	\sum_{i = 1}^{L}\sum_{j = 1}^{L} t_{1, i}\binom{L}{\left\lvert i - j \right\rvert} \, x^{j} = x^k.
\end{equation}
Comparing the powers of $x$ in the two sides leads to a non-homogeneous system that must be solved to determine the coefficients $t_{1,i}$,
\begin{equation} \label{eq:system}
	\sum_{i = 1}^{L} t_{1,i} \binom{L}{\left\lvert i - j \right\rvert} \bmod{2} = \begin{cases}
		1 & \text{if} \quad j = k \\
		0 & \text{otherwise}.
	\end{cases}  
\end{equation}
Writing this equation in a matrix form with matrix elements $M_{ij} = \binom{L}{\left\lvert i - j\right\rvert} \bmod{2}$, then this system has a unique nontrivial solution if and only if $\det M \neq 0$. The matrices that we find in this case are $L \times L$ circulant matrices with elements $\left\{ 0, 1\right\}$. It is straightforward to compute the determinant numerically and find the solution of the system. Examples are shown in Fig. \ref{membraneoperator}.

Let us denote a configuration generated by $S_{1}(x)$ that satisfies Eq. \eqref{eq:system} as $\tilde{\mathcal{M}}_{i}^{a}$, where $i$ specifies the position of the isolated $X$-operator in the sublattice $a$. The tilde over $\mathcal{M}$ is to point out that the corresponding operators cannot be closed. Stacking this product along the $z$-direction leads to
\begin{eqnarray} \label{WLA1}
	\prod_{z} \prod_{p \, \in \, \tilde{\mathcal{M}}_{i}^{a}} \mathcal{O}_{p} = \prod_{p \, \in \, \mathcal{L}^{a}_{i}} X_{p} \equiv W_{i}^{a},
\end{eqnarray}
which is the Wilson line (see Eq. (8) of the main text). The Wilson line are symmetries of the system independent of the sizes. However, they are genuine symmetries only for sizes of the form $L = k (2^n - 2^m)$, with $n > m + 1$, whereas for the remaining sizes they can always be expressed as a product of TD operators.

For system sizes $L = k (2^n - 2^m)$ there is a simple way to understand that the Wilson lines cannot be reduced to a product of TD operators. In this case, there are closed membrane operators $M_{I}^{a}$, which satisfy the nontrivial commutation of Eq. \eqref{projective}. If $W_{J}^{b}$ is a product of TD operators, then $M_{I}^{a}$ should not commute with one $\mathcal{O}_{i}$ and, consequently, $M_{I}^{a}$ will not commute with the Hamiltonian. Therefore, if there are fractal membrane operators that act as symmetries, the Wilson lines cannot be expressed as a product of TD operators.

%% file: CCmodel_Letter_final-Heitor.bbl
\begin{thebibliography}{60}%
\makeatletter
\providecommand \@ifxundefined [1]{%
 \@ifx{#1\undefined}
}%
\providecommand \@ifnum [1]{%
 \ifnum #1\expandafter \@firstoftwo
 \else \expandafter \@secondoftwo
 \fi
}%
\providecommand \@ifx [1]{%
 \ifx #1\expandafter \@firstoftwo
 \else \expandafter \@secondoftwo
 \fi
}%
\providecommand \natexlab [1]{#1}%
\providecommand \enquote  [1]{``#1''}%
\providecommand \bibnamefont  [1]{#1}%
\providecommand \bibfnamefont [1]{#1}%
\providecommand \citenamefont [1]{#1}%
\providecommand \href@noop [0]{\@secondoftwo}%
\providecommand \href [0]{\begingroup \@sanitize@url \@href}%
\providecommand \@href[1]{\@@startlink{#1}\@@href}%
\providecommand \@@href[1]{\endgroup#1\@@endlink}%
\providecommand \@sanitize@url [0]{\catcode `\\12\catcode `\$12\catcode
  `\&12\catcode `\#12\catcode `\^12\catcode `\_12\catcode `\%12\relax}%
\providecommand \@@startlink[1]{}%
\providecommand \@@endlink[0]{}%
\providecommand \url  [0]{\begingroup\@sanitize@url \@url }%
\providecommand \@url [1]{\endgroup\@href {#1}{\urlprefix }}%
\providecommand \urlprefix  [0]{URL }%
\providecommand \Eprint [0]{\href }%
\providecommand \doibase [0]{https://doi.org/}%
\providecommand \selectlanguage [0]{\@gobble}%
\providecommand \bibinfo  [0]{\@secondoftwo}%
\providecommand \bibfield  [0]{\@secondoftwo}%
\providecommand \translation [1]{[#1]}%
\providecommand \BibitemOpen [0]{}%
\providecommand \bibitemStop [0]{}%
\providecommand \bibitemNoStop [0]{.\EOS\space}%
\providecommand \EOS [0]{\spacefactor3000\relax}%
\providecommand \BibitemShut  [1]{\csname bibitem#1\endcsname}%
\let\auto@bib@innerbib\@empty
\bibitem [{\citenamefont {Wen}(2017)}]{Wen_2017}%
  \BibitemOpen
  \bibfield  {author} {\bibinfo {author} {\bibfnamefont {X.-G.}\ \bibnamefont
  {Wen}},\ }\bibfield  {journal} {\bibinfo  {journal} {Reviews of Modern
  Physics}\ }\textbf {\bibinfo {volume} {89}},\ \href
  {https://doi.org/10.1103/RevModPhys.89.041004} {10.1103/RevModPhys.89.041004}
  (\bibinfo {year} {2017}),\ \Eprint {https://arxiv.org/abs/1610.03911}
  {arXiv:1610.03911 [cond-mat.str-el]} \BibitemShut {NoStop}%
\bibitem [{\citenamefont {Chamon}(2005)}]{Chamon_2005}%
  \BibitemOpen
  \bibfield  {author} {\bibinfo {author} {\bibfnamefont {C.}~\bibnamefont
  {Chamon}},\ }\bibfield  {journal} {\bibinfo  {journal} {Physical Review
  Letters}\ }\textbf {\bibinfo {volume} {94}},\ \href
  {https://doi.org/10.1103/PhysRevLett.94.040402}
  {10.1103/PhysRevLett.94.040402} (\bibinfo {year} {2005}),\ \Eprint
  {https://arxiv.org/abs/cond-mat/0404182} {arXiv:cond-mat/0404182
  [cond-mat.str-el]} \BibitemShut {NoStop}%
\bibitem [{\citenamefont {Bravyi}\ \emph {et~al.}(2011)\citenamefont {Bravyi},
  \citenamefont {Leemhuis},\ and\ \citenamefont {Terhal}}]{Bravyi_2011}%
  \BibitemOpen
  \bibfield  {author} {\bibinfo {author} {\bibfnamefont {S.}~\bibnamefont
  {Bravyi}}, \bibinfo {author} {\bibfnamefont {B.}~\bibnamefont {Leemhuis}},\
  and\ \bibinfo {author} {\bibfnamefont {B.~M.}\ \bibnamefont {Terhal}},\
  }\href {https://doi.org/10.1016/j.aop.2010.11.002} {\bibfield  {journal}
  {\bibinfo  {journal} {Annals of Physics}\ }\textbf {\bibinfo {volume}
  {326}},\ \bibinfo {pages} {839} (\bibinfo {year} {2011})},\ \Eprint
  {https://arxiv.org/abs/1006.4871} {arXiv:1006.4871 [quant-ph]} \BibitemShut
  {NoStop}%
\bibitem [{\citenamefont {Haah}(2011)}]{Haah_2011}%
  \BibitemOpen
  \bibfield  {author} {\bibinfo {author} {\bibfnamefont {J.}~\bibnamefont
  {Haah}},\ }\bibfield  {journal} {\bibinfo  {journal} {Physical Review A}\
  }\textbf {\bibinfo {volume} {83}},\ \href
  {https://doi.org/10.1103/PhysRevA.83.042330} {10.1103/PhysRevA.83.042330}
  (\bibinfo {year} {2011}),\ \Eprint {https://arxiv.org/abs/1101.1962}
  {arXiv:1101.1962 [quant-ph]} \BibitemShut {NoStop}%
\bibitem [{\citenamefont {Vijay}\ \emph {et~al.}(2015)\citenamefont {Vijay},
  \citenamefont {Haah},\ and\ \citenamefont {Fu}}]{Vijay:2015mka}%
  \BibitemOpen
  \bibfield  {author} {\bibinfo {author} {\bibfnamefont {S.}~\bibnamefont
  {Vijay}}, \bibinfo {author} {\bibfnamefont {J.}~\bibnamefont {Haah}},\ and\
  \bibinfo {author} {\bibfnamefont {L.}~\bibnamefont {Fu}},\ }\href
  {https://doi.org/10.1103/PhysRevB.92.235136} {\bibfield  {journal} {\bibinfo
  {journal} {Physical Review B}\ }\textbf {\bibinfo {volume} {92}},\ \bibinfo
  {pages} {235136} (\bibinfo {year} {2015})},\ \Eprint
  {https://arxiv.org/abs/1505.02576} {arXiv:1505.02576 [cond-mat.str-el]}
  \BibitemShut {NoStop}%
\bibitem [{\citenamefont {Prem}\ \emph {et~al.}(2017)\citenamefont {Prem},
  \citenamefont {Haah},\ and\ \citenamefont {Nandkishore}}]{Nandkishore2017}%
  \BibitemOpen
  \bibfield  {author} {\bibinfo {author} {\bibfnamefont {A.}~\bibnamefont
  {Prem}}, \bibinfo {author} {\bibfnamefont {J.}~\bibnamefont {Haah}},\ and\
  \bibinfo {author} {\bibfnamefont {R.}~\bibnamefont {Nandkishore}},\
  }\bibfield  {journal} {\bibinfo  {journal} {Physical Review B}\ }\textbf
  {\bibinfo {volume} {95}},\ \href {https://doi.org/10.1103/PhysRevB.95.155133}
  {10.1103/PhysRevB.95.155133} (\bibinfo {year} {2017}),\ \Eprint
  {https://arxiv.org/abs/1702.02952} {arXiv:1702.02952 [cond-mat.stat-mech]}
  \BibitemShut {NoStop}%
\bibitem [{\citenamefont {Slagle}\ and\ \citenamefont
  {Kim}(2017)}]{SlagleKim2017}%
  \BibitemOpen
  \bibfield  {author} {\bibinfo {author} {\bibfnamefont {K.}~\bibnamefont
  {Slagle}}\ and\ \bibinfo {author} {\bibfnamefont {Y.~B.}\ \bibnamefont
  {Kim}},\ }\bibfield  {journal} {\bibinfo  {journal} {Physical Review B}\
  }\textbf {\bibinfo {volume} {96}},\ \href
  {https://doi.org/10.1103/PhysRevB.96.195139} {10.1103/PhysRevB.96.195139}
  (\bibinfo {year} {2017}),\ \Eprint {https://arxiv.org/abs/1708.04619}
  {arXiv:1708.04619 [cond-mat.str-el]} \BibitemShut {NoStop}%
\bibitem [{\citenamefont {You}\ \emph {et~al.}(2020{\natexlab{a}})\citenamefont
  {You}, \citenamefont {Devakul}, \citenamefont {Burnell},\ and\ \citenamefont
  {Sondhi}}]{You:2018zhj}%
  \BibitemOpen
  \bibfield  {author} {\bibinfo {author} {\bibfnamefont {Y.}~\bibnamefont
  {You}}, \bibinfo {author} {\bibfnamefont {T.}~\bibnamefont {Devakul}},
  \bibinfo {author} {\bibfnamefont {F.~J.}\ \bibnamefont {Burnell}},\ and\
  \bibinfo {author} {\bibfnamefont {S.~L.}\ \bibnamefont {Sondhi}},\ }\href
  {https://doi.org/10.1016/j.aop.2020.168140} {\bibfield  {journal} {\bibinfo
  {journal} {Annals of Physics}\ }\textbf {\bibinfo {volume} {416}},\ \bibinfo
  {pages} {168140} (\bibinfo {year} {2020}{\natexlab{a}})},\ \Eprint
  {https://arxiv.org/abs/1805.09800} {arXiv:1805.09800 [cond-mat.str-el]}
  \BibitemShut {NoStop}%
\bibitem [{\citenamefont {Shirley}\ \emph {et~al.}(2019)\citenamefont
  {Shirley}, \citenamefont {Slagle},\ and\ \citenamefont {Chen}}]{Shirley2019}%
  \BibitemOpen
  \bibfield  {author} {\bibinfo {author} {\bibfnamefont {W.}~\bibnamefont
  {Shirley}}, \bibinfo {author} {\bibfnamefont {K.}~\bibnamefont {Slagle}},\
  and\ \bibinfo {author} {\bibfnamefont {X.}~\bibnamefont {Chen}},\ }\bibfield
  {journal} {\bibinfo  {journal} {{SciPost} Physics}\ }\textbf {\bibinfo
  {volume} {6}},\ \href {https://doi.org/10.21468/SciPostPhys.6.1.015}
  {10.21468/SciPostPhys.6.1.015} (\bibinfo {year} {2019}),\ \Eprint
  {https://arxiv.org/abs/1803.10426} {arXiv:1803.10426 [cond-mat.str-el]}
  \BibitemShut {NoStop}%
\bibitem [{\citenamefont {Shirley}\ \emph {et~al.}(2020)\citenamefont
  {Shirley}, \citenamefont {Slagle},\ and\ \citenamefont {Chen}}]{Chen2019a}%
  \BibitemOpen
  \bibfield  {author} {\bibinfo {author} {\bibfnamefont {W.}~\bibnamefont
  {Shirley}}, \bibinfo {author} {\bibfnamefont {K.}~\bibnamefont {Slagle}},\
  and\ \bibinfo {author} {\bibfnamefont {X.}~\bibnamefont {Chen}},\ }\bibfield
  {journal} {\bibinfo  {journal} {Physical Review B}\ }\textbf {\bibinfo
  {volume} {102}},\ \href {https://doi.org/10.1103/PhysRevB.102.115103}
  {10.1103/PhysRevB.102.115103} (\bibinfo {year} {2020}),\ \Eprint
  {https://arxiv.org/abs/1907.09048} {arXiv:1907.09048 [cond-mat.str-el]}
  \BibitemShut {NoStop}%
\bibitem [{\citenamefont {Fuji}(2019)}]{Fuji2019}%
  \BibitemOpen
  \bibfield  {author} {\bibinfo {author} {\bibfnamefont {Y.}~\bibnamefont
  {Fuji}},\ }\bibfield  {journal} {\bibinfo  {journal} {Physical Review B}\
  }\textbf {\bibinfo {volume} {100}},\ \href
  {https://doi.org/10.1103/PhysRevB.100.235115} {10.1103/PhysRevB.100.235115}
  (\bibinfo {year} {2019}),\ \Eprint {https://arxiv.org/abs/1908.02257}
  {arXiv:1908.02257 [cond-mat.str-el]} \BibitemShut {NoStop}%
\bibitem [{\citenamefont {Fontana}\ \emph {et~al.}(2021)\citenamefont
  {Fontana}, \citenamefont {Gomes},\ and\ \citenamefont
  {Chamon}}]{Fontana:2020tby}%
  \BibitemOpen
  \bibfield  {author} {\bibinfo {author} {\bibfnamefont {W.}~\bibnamefont
  {Fontana}}, \bibinfo {author} {\bibfnamefont {P.}~\bibnamefont {Gomes}},\
  and\ \bibinfo {author} {\bibfnamefont {C.}~\bibnamefont {Chamon}},\ }\href
  {https://doi.org/10.21468/SciPostPhysCore.4.2.012} {\bibfield  {journal}
  {\bibinfo  {journal} {{SciPost} Physics Core}\ }\textbf {\bibinfo {volume}
  {4}},\ \bibinfo {pages} {012} (\bibinfo {year} {2021})},\ \Eprint
  {https://arxiv.org/abs/2006.10071} {arXiv:2006.10071 [cond-mat.str-el]}
  \BibitemShut {NoStop}%
\bibitem [{\citenamefont {Fontana}\ \emph {et~al.}(2022)\citenamefont
  {Fontana}, \citenamefont {Gomes},\ and\ \citenamefont
  {Chamon}}]{Fontana:2021zwt}%
  \BibitemOpen
  \bibfield  {author} {\bibinfo {author} {\bibfnamefont {W.}~\bibnamefont
  {Fontana}}, \bibinfo {author} {\bibfnamefont {P.}~\bibnamefont {Gomes}},\
  and\ \bibinfo {author} {\bibfnamefont {C.}~\bibnamefont {Chamon}},\ }\href
  {https://doi.org/10.21468/SciPostPhys.12.2.064} {\bibfield  {journal}
  {\bibinfo  {journal} {{SciPost} Physics}\ }\textbf {\bibinfo {volume} {12}},\
  \bibinfo {pages} {064} (\bibinfo {year} {2022})},\ \Eprint
  {https://arxiv.org/abs/2103.02713} {arXiv:2103.02713 [cond-mat.str-el]}
  \BibitemShut {NoStop}%
\bibitem [{\citenamefont {Nandkishore}\ and\ \citenamefont
  {Hermele}(2019)}]{Nandkishore:2018sel}%
  \BibitemOpen
  \bibfield  {author} {\bibinfo {author} {\bibfnamefont {R.~M.}\ \bibnamefont
  {Nandkishore}}\ and\ \bibinfo {author} {\bibfnamefont {M.}~\bibnamefont
  {Hermele}},\ }\href
  {https://doi.org/10.1146/annurev-conmatphys-031218-013604} {\bibfield
  {journal} {\bibinfo  {journal} {Annual Review of Condensed Matter Physics}\
  }\textbf {\bibinfo {volume} {10}},\ \bibinfo {pages} {295} (\bibinfo {year}
  {2019})},\ \Eprint {https://arxiv.org/abs/1803.11196} {arXiv:1803.11196
  [cond-mat.str-el]} \BibitemShut {NoStop}%
\bibitem [{\citenamefont {Pretko}\ \emph {et~al.}(2020)\citenamefont {Pretko},
  \citenamefont {Chen},\ and\ \citenamefont {You}}]{Pretko:2020cko}%
  \BibitemOpen
  \bibfield  {author} {\bibinfo {author} {\bibfnamefont {M.}~\bibnamefont
  {Pretko}}, \bibinfo {author} {\bibfnamefont {X.}~\bibnamefont {Chen}},\ and\
  \bibinfo {author} {\bibfnamefont {Y.}~\bibnamefont {You}},\ }\href
  {https://doi.org/10.1142/S0217751X20300033} {\bibfield  {journal} {\bibinfo
  {journal} {International Journal of Modern Physics A}\ }\textbf {\bibinfo
  {volume} {35}},\ \bibinfo {pages} {2030003} (\bibinfo {year} {2020})},\
  \Eprint {https://arxiv.org/abs/2001.01722} {arXiv:2001.01722
  [cond-mat.str-el]} \BibitemShut {NoStop}%
\bibitem [{\citenamefont {Vijay}\ \emph {et~al.}(2016)\citenamefont {Vijay},
  \citenamefont {Haah},\ and\ \citenamefont {Fu}}]{Vijay:2016phm}%
  \BibitemOpen
  \bibfield  {author} {\bibinfo {author} {\bibfnamefont {S.}~\bibnamefont
  {Vijay}}, \bibinfo {author} {\bibfnamefont {J.}~\bibnamefont {Haah}},\ and\
  \bibinfo {author} {\bibfnamefont {L.}~\bibnamefont {Fu}},\ }\bibfield
  {journal} {\bibinfo  {journal} {Physical Review B}\ }\textbf {\bibinfo
  {volume} {94}},\ \href {https://doi.org/10.1103/PhysRevB.94.235157}
  {10.1103/PhysRevB.94.235157} (\bibinfo {year} {2016}),\ \Eprint
  {https://arxiv.org/abs/1603.04442} {arXiv:1603.04442 [cond-mat.str-el]}
  \BibitemShut {NoStop}%
\bibitem [{\citenamefont {Pretko}(2017)}]{Pretko2017}%
  \BibitemOpen
  \bibfield  {author} {\bibinfo {author} {\bibfnamefont {M.}~\bibnamefont
  {Pretko}},\ }\bibfield  {journal} {\bibinfo  {journal} {Physical Review B}\
  }\textbf {\bibinfo {volume} {95}},\ \href
  {https://doi.org/10.1103/PhysRevB.95.115139} {10.1103/PhysRevB.95.115139}
  (\bibinfo {year} {2017}),\ \Eprint {https://arxiv.org/abs/1604.05329}
  {arXiv:1604.05329 [cond-mat.str-el]} \BibitemShut {NoStop}%
\bibitem [{\citenamefont {Jangjan}\ and\ \citenamefont
  {Hosseini}(2022)}]{Jangjan2022}%
  \BibitemOpen
  \bibfield  {author} {\bibinfo {author} {\bibfnamefont {M.}~\bibnamefont
  {Jangjan}}\ and\ \bibinfo {author} {\bibfnamefont {M.~V.}\ \bibnamefont
  {Hosseini}},\ }\bibfield  {journal} {\bibinfo  {journal} {Physical Review B}\
  }\textbf {\bibinfo {volume} {106}},\ \href
  {https://doi.org/10.1103/PhysRevB.106.205111} {10.1103/PhysRevB.106.205111}
  (\bibinfo {year} {2022}),\ \Eprint {https://arxiv.org/abs/2203.13160}
  {arXiv:2203.13160 [cond-mat.mes-hall]} \BibitemShut {NoStop}%
\bibitem [{\citenamefont {Delfino}\ and\ \citenamefont
  {You}(2023)}]{Delfino2023}%
  \BibitemOpen
  \bibfield  {author} {\bibinfo {author} {\bibfnamefont {G.}~\bibnamefont
  {Delfino}}\ and\ \bibinfo {author} {\bibfnamefont {Y.}~\bibnamefont {You}},\
  }\bibfield  {journal} {\bibinfo  {journal} {arXiv e-prints}\ }\href
  {https://doi.org/10.48550/arXiv.2310.09490} {10.48550/arXiv.2310.09490}
  (\bibinfo {year} {2023}),\ \Eprint {https://arxiv.org/abs/2310.09490}
  {arXiv:2310.09490 [cond-mat.str-el]} \BibitemShut {NoStop}%
\bibitem [{\citenamefont {Gaiotto}\ \emph {et~al.}(2015)\citenamefont
  {Gaiotto}, \citenamefont {Kapustin}, \citenamefont {Seiberg},\ and\
  \citenamefont {Willett}}]{Gaiotto:2014kfa}%
  \BibitemOpen
  \bibfield  {author} {\bibinfo {author} {\bibfnamefont {D.}~\bibnamefont
  {Gaiotto}}, \bibinfo {author} {\bibfnamefont {A.}~\bibnamefont {Kapustin}},
  \bibinfo {author} {\bibfnamefont {N.}~\bibnamefont {Seiberg}},\ and\ \bibinfo
  {author} {\bibfnamefont {B.}~\bibnamefont {Willett}},\ }\href
  {https://doi.org/10.1007/jhep02(2015)172} {\bibfield  {journal} {\bibinfo
  {journal} {Journal of High Energy Physics}\ }\textbf {\bibinfo {volume}
  {2015}},\ \bibinfo {pages} {172} (\bibinfo {year} {2015})},\ \Eprint
  {https://arxiv.org/abs/1412.5148} {arXiv:1412.5148 [hep-th]} \BibitemShut
  {NoStop}%
\bibitem [{\citenamefont {Gomes}(2023)}]{Gomes:2023ahz}%
  \BibitemOpen
  \bibfield  {author} {\bibinfo {author} {\bibfnamefont {P.~R.~S.}\
  \bibnamefont {Gomes}},\ }\bibfield  {journal} {\bibinfo  {journal} {{SciPost}
  Physics Lecture Notes}\ }\textbf {\bibinfo {volume} {74}},\ \href
  {https://doi.org/10.21468/SciPostPhysLectNotes.74}
  {10.21468/SciPostPhysLectNotes.74} (\bibinfo {year} {2023}),\ \Eprint
  {https://arxiv.org/abs/2303.01817} {arXiv:2303.01817 [hep-th]} \BibitemShut
  {NoStop}%
\bibitem [{\citenamefont {Brennan}\ and\ \citenamefont
  {Hong}(2023)}]{Brennan:2023mmt}%
  \BibitemOpen
  \bibfield  {author} {\bibinfo {author} {\bibfnamefont {T.~D.}\ \bibnamefont
  {Brennan}}\ and\ \bibinfo {author} {\bibfnamefont {S.}~\bibnamefont {Hong}},\
  }\bibfield  {journal} {\bibinfo  {journal} {arXiv e-prints}\ }\href
  {https://doi.org/10.48550/arXiv.2306.00912} {10.48550/arXiv.2306.00912}
  (\bibinfo {year} {2023}),\ \Eprint {https://arxiv.org/abs/2306.00912}
  {arXiv:2306.00912 [hep-ph]} \BibitemShut {NoStop}%
\bibitem [{\citenamefont {Luo}\ \emph {et~al.}(2023)\citenamefont {Luo},
  \citenamefont {Wang},\ and\ \citenamefont {Wang}}]{Luo:2023ive}%
  \BibitemOpen
  \bibfield  {author} {\bibinfo {author} {\bibfnamefont {R.}~\bibnamefont
  {Luo}}, \bibinfo {author} {\bibfnamefont {Q.-R.}\ \bibnamefont {Wang}},\ and\
  \bibinfo {author} {\bibfnamefont {Y.-N.}\ \bibnamefont {Wang}},\ }\bibfield
  {journal} {\bibinfo  {journal} {arXiv e-prints}\ }\href
  {https://doi.org/10.48550/arXiv.2307.09215} {10.48550/arXiv.2307.09215}
  (\bibinfo {year} {2023}),\ \Eprint {https://arxiv.org/abs/2307.09215}
  {arXiv:2307.09215 [hep-th]} \BibitemShut {NoStop}%
\bibitem [{\citenamefont {Bhardwaj}\ \emph {et~al.}(2023)\citenamefont
  {Bhardwaj}, \citenamefont {Bottini}, \citenamefont {Fraser-Taliente},
  \citenamefont {Gladden}, \citenamefont {Gould}, \citenamefont {Platschorre},\
  and\ \citenamefont {Tillim}}]{Bhardwaj:2023kri}%
  \BibitemOpen
  \bibfield  {author} {\bibinfo {author} {\bibfnamefont {L.}~\bibnamefont
  {Bhardwaj}}, \bibinfo {author} {\bibfnamefont {L.~E.}\ \bibnamefont
  {Bottini}}, \bibinfo {author} {\bibfnamefont {L.}~\bibnamefont
  {Fraser-Taliente}}, \bibinfo {author} {\bibfnamefont {L.}~\bibnamefont
  {Gladden}}, \bibinfo {author} {\bibfnamefont {D.~S.~W.}\ \bibnamefont
  {Gould}}, \bibinfo {author} {\bibfnamefont {A.}~\bibnamefont {Platschorre}},\
  and\ \bibinfo {author} {\bibfnamefont {H.}~\bibnamefont {Tillim}},\
  }\bibfield  {journal} {\bibinfo  {journal} {arXiv e-prints}\ }\href
  {https://doi.org/10.48550/arXiv.2307.07547} {10.48550/arXiv.2307.07547}
  (\bibinfo {year} {2023}),\ \Eprint {https://arxiv.org/abs/2307.07547}
  {arXiv:2307.07547 [hep-th]} \BibitemShut {NoStop}%
\bibitem [{\citenamefont {Wen}(2019)}]{Wen_2019}%
  \BibitemOpen
  \bibfield  {author} {\bibinfo {author} {\bibfnamefont {X.-G.}\ \bibnamefont
  {Wen}},\ }\bibfield  {journal} {\bibinfo  {journal} {Physical Review B}\
  }\textbf {\bibinfo {volume} {99}},\ \href
  {https://doi.org/10.1103/PhysRevB.99.205139} {10.1103/PhysRevB.99.205139}
  (\bibinfo {year} {2019}),\ \Eprint {https://arxiv.org/abs/1812.02517}
  {arXiv:1812.02517 [cond-mat.str-el]} \BibitemShut {NoStop}%
\bibitem [{\citenamefont {Qi}\ \emph {et~al.}(2021)\citenamefont {Qi},
  \citenamefont {Radzihovsky},\ and\ \citenamefont {Hermele}}]{Qi:2020jrf}%
  \BibitemOpen
  \bibfield  {author} {\bibinfo {author} {\bibfnamefont {M.}~\bibnamefont
  {Qi}}, \bibinfo {author} {\bibfnamefont {L.}~\bibnamefont {Radzihovsky}},\
  and\ \bibinfo {author} {\bibfnamefont {M.}~\bibnamefont {Hermele}},\ }\href
  {https://doi.org/10.1016/j.aop.2020.168360} {\bibfield  {journal} {\bibinfo
  {journal} {Annals of Physics}\ }\textbf {\bibinfo {volume} {424}},\ \bibinfo
  {pages} {168360} (\bibinfo {year} {2021})},\ \Eprint
  {https://arxiv.org/abs/2010.02254} {arXiv:2010.02254 [cond-mat.str-el]}
  \BibitemShut {NoStop}%
\bibitem [{\citenamefont {Rayhaun}\ and\ \citenamefont
  {Williamson}(2023)}]{Rayhaun:2021ocs}%
  \BibitemOpen
  \bibfield  {author} {\bibinfo {author} {\bibfnamefont {B.~C.}\ \bibnamefont
  {Rayhaun}}\ and\ \bibinfo {author} {\bibfnamefont {D.~J.}\ \bibnamefont
  {Williamson}},\ }\bibfield  {journal} {\bibinfo  {journal} {{SciPost}
  Physics}\ }\textbf {\bibinfo {volume} {15}},\ \href
  {https://doi.org/10.21468/SciPostPhys.15.1.017}
  {10.21468/SciPostPhys.15.1.017} (\bibinfo {year} {2023}),\ \Eprint
  {https://arxiv.org/abs/2112.12735} {arXiv:2112.12735 [cond-mat.str-el]}
  \BibitemShut {NoStop}%
\bibitem [{\citenamefont {Bonderson}\ and\ \citenamefont
  {Nayak}(2013)}]{Bonderson_2013}%
  \BibitemOpen
  \bibfield  {author} {\bibinfo {author} {\bibfnamefont {P.}~\bibnamefont
  {Bonderson}}\ and\ \bibinfo {author} {\bibfnamefont {C.}~\bibnamefont
  {Nayak}},\ }\bibfield  {journal} {\bibinfo  {journal} {Physical Review B}\
  }\textbf {\bibinfo {volume} {87}},\ \href
  {https://doi.org/10.1103/PhysRevB.87.195451} {10.1103/PhysRevB.87.195451}
  (\bibinfo {year} {2013}),\ \Eprint {https://arxiv.org/abs/1212.6395}
  {arXiv:1212.6395 [cond-mat.str-el]} \BibitemShut {NoStop}%
\bibitem [{\citenamefont {Huxford}\ \emph {et~al.}(2023)\citenamefont
  {Huxford}, \citenamefont {Nguyen},\ and\ \citenamefont
  {Kim}}]{Huxford:2023bhb}%
  \BibitemOpen
  \bibfield  {author} {\bibinfo {author} {\bibfnamefont {J.}~\bibnamefont
  {Huxford}}, \bibinfo {author} {\bibfnamefont {D.~X.}\ \bibnamefont
  {Nguyen}},\ and\ \bibinfo {author} {\bibfnamefont {Y.~B.}\ \bibnamefont
  {Kim}},\ }\bibfield  {journal} {\bibinfo  {journal} {arXiv e-prints}\ }\href
  {https://doi.org/10.48550/arXiv.2305.07063} {10.48550/arXiv.2305.07063}
  (\bibinfo {year} {2023}),\ \Eprint {https://arxiv.org/abs/2305.07063}
  {arXiv:2305.07063 [cond-mat.str-el]} \BibitemShut {NoStop}%
\bibitem [{\citenamefont {McGreevy}(2023)}]{McGreevy_2023}%
  \BibitemOpen
  \bibfield  {author} {\bibinfo {author} {\bibfnamefont {J.}~\bibnamefont
  {McGreevy}},\ }\href
  {https://doi.org/10.1146/annurev-conmatphys-040721-021029} {\bibfield
  {journal} {\bibinfo  {journal} {Annual Review of Condensed Matter Physics}\
  }\textbf {\bibinfo {volume} {14}},\ \bibinfo {pages} {57} (\bibinfo {year}
  {2023})},\ \Eprint {https://arxiv.org/abs/2204.03045} {arXiv:2204.03045
  [cond-mat.str-el]} \BibitemShut {NoStop}%
\bibitem [{\citenamefont {Pace}\ and\ \citenamefont
  {Wen}(2023)}]{Pace_2023emergent}%
  \BibitemOpen
  \bibfield  {author} {\bibinfo {author} {\bibfnamefont {S.~D.}\ \bibnamefont
  {Pace}}\ and\ \bibinfo {author} {\bibfnamefont {X.-G.}\ \bibnamefont {Wen}},\
  }\bibfield  {journal} {\bibinfo  {journal} {Physical Review B}\ }\textbf
  {\bibinfo {volume} {108}},\ \href
  {https://doi.org/10.1103/PhysRevB.108.195147} {10.1103/PhysRevB.108.195147}
  (\bibinfo {year} {2023}),\ \Eprint {https://arxiv.org/abs/2301.05261}
  {arXiv:2301.05261 [cond-mat.str-el]} \BibitemShut {NoStop}%
\bibitem [{\citenamefont {Cherman}\ and\ \citenamefont
  {Jacobson}(2023)}]{cherman2023emergent}%
  \BibitemOpen
  \bibfield  {author} {\bibinfo {author} {\bibfnamefont {A.}~\bibnamefont
  {Cherman}}\ and\ \bibinfo {author} {\bibfnamefont {T.}~\bibnamefont
  {Jacobson}},\ }\bibfield  {journal} {\bibinfo  {journal} {arXiv e-prints}\
  }\href {https://doi.org/10.48550/arXiv.2304.13751}
  {10.48550/arXiv.2304.13751} (\bibinfo {year} {2023}),\ \Eprint
  {https://arxiv.org/abs/2304.13751} {arXiv:2304.13751 [hep-th]} \BibitemShut
  {NoStop}%
\bibitem [{\citenamefont {Fröhlich}\ \emph {et~al.}(2004)\citenamefont
  {Fröhlich}, \citenamefont {Fuchs}, \citenamefont {Runkel},\ and\
  \citenamefont {Schweigert}}]{Froehlich2004}%
  \BibitemOpen
  \bibfield  {author} {\bibinfo {author} {\bibfnamefont {J.}~\bibnamefont
  {Fröhlich}}, \bibinfo {author} {\bibfnamefont {J.}~\bibnamefont {Fuchs}},
  \bibinfo {author} {\bibfnamefont {I.}~\bibnamefont {Runkel}},\ and\ \bibinfo
  {author} {\bibfnamefont {C.}~\bibnamefont {Schweigert}},\ }\bibfield
  {journal} {\bibinfo  {journal} {Physical Review Letters}\ }\textbf {\bibinfo
  {volume} {93}},\ \href {https://doi.org/10.1103/PhysRevLett.93.070601}
  {10.1103/PhysRevLett.93.070601} (\bibinfo {year} {2004}),\ \Eprint
  {https://arxiv.org/abs/cond-mat/0404051} {arXiv:cond-mat/0404051
  [cond-mat.stat-mech]} \BibitemShut {NoStop}%
\bibitem [{\citenamefont {Fröhlich}\ \emph {et~al.}(2010)\citenamefont
  {Fröhlich}, \citenamefont {Fuchs}, \citenamefont {Runkel},\ and\
  \citenamefont {Schweigert}}]{FROeHLICH2010}%
  \BibitemOpen
  \bibfield  {author} {\bibinfo {author} {\bibfnamefont {J.}~\bibnamefont
  {Fröhlich}}, \bibinfo {author} {\bibfnamefont {J.}~\bibnamefont {Fuchs}},
  \bibinfo {author} {\bibfnamefont {I.}~\bibnamefont {Runkel}},\ and\ \bibinfo
  {author} {\bibfnamefont {C.}~\bibnamefont {Schweigert}},\ }\bibinfo {title}
  {Defect lines, dualities, and generalised orbifolds},\ in\ \href
  {https://doi.org/10.1142/9789814304634\_0056} {\emph {\bibinfo {booktitle}
  {XVIth International Congress on Mathematical Physics}}}\ (\bibinfo
  {publisher} {WORLD SCIENTIFIC},\ \bibinfo {address} {Prague, Czech
  Republic},\ \bibinfo {year} {2010})\ pp.\ \bibinfo {pages} {608--613},\
  \Eprint {https://arxiv.org/abs/0909.5013} {arXiv:0909.5013 [math-ph]}
  \BibitemShut {NoStop}%
\bibitem [{\citenamefont {Chang}\ \emph {et~al.}(2019)\citenamefont {Chang},
  \citenamefont {Lin}, \citenamefont {Shao}, \citenamefont {Wang},\ and\
  \citenamefont {Yin}}]{Chang2019}%
  \BibitemOpen
  \bibfield  {author} {\bibinfo {author} {\bibfnamefont {C.-M.}\ \bibnamefont
  {Chang}}, \bibinfo {author} {\bibfnamefont {Y.-H.}\ \bibnamefont {Lin}},
  \bibinfo {author} {\bibfnamefont {S.-H.}\ \bibnamefont {Shao}}, \bibinfo
  {author} {\bibfnamefont {Y.}~\bibnamefont {Wang}},\ and\ \bibinfo {author}
  {\bibfnamefont {X.}~\bibnamefont {Yin}},\ }\bibfield  {journal} {\bibinfo
  {journal} {Journal of High Energy Physics}\ }\textbf {\bibinfo {volume}
  {2019}},\ \href {https://doi.org/10.1007/jhep01(2019)026}
  {10.1007/jhep01(2019)026} (\bibinfo {year} {2019}),\ \Eprint
  {https://arxiv.org/abs/1802.04445} {arXiv:1802.04445 [hep-th]} \BibitemShut
  {NoStop}%
\bibitem [{\citenamefont {Schafer-Nameki}(2024)}]{SchaferNameki2023}%
  \BibitemOpen
  \bibfield  {author} {\bibinfo {author} {\bibfnamefont {S.}~\bibnamefont
  {Schafer-Nameki}},\ }\href {https://doi.org/10.1016/j.physrep.2024.01.007}
  {\bibfield  {journal} {\bibinfo  {journal} {Physics Reports}\ }\textbf
  {\bibinfo {volume} {1063}},\ \bibinfo {pages} {1} (\bibinfo {year} {2024})},\
  \Eprint {https://arxiv.org/abs/2305.18296} {arXiv:2305.18296 [hep-th]}
  \BibitemShut {NoStop}%
\bibitem [{\citenamefont {Shao}(2023)}]{Shao2023}%
  \BibitemOpen
  \bibfield  {author} {\bibinfo {author} {\bibfnamefont {S.-H.}\ \bibnamefont
  {Shao}},\ }\bibfield  {journal} {\bibinfo  {journal} {arXiv e-prints}\ }\href
  {https://doi.org/10.48550/ARXIV.2308.00747} {10.48550/ARXIV.2308.00747}
  (\bibinfo {year} {2023}),\ \Eprint {https://arxiv.org/abs/2308.00747}
  {arXiv:2308.00747 [hep-th]} \BibitemShut {NoStop}%
\bibitem [{\citenamefont {'t~Hooft}(1980)}]{tHooft:1979rat}%
  \BibitemOpen
  \bibfield  {author} {\bibinfo {author} {\bibfnamefont {G.}~\bibnamefont
  {'t~Hooft}},\ }\bibinfo {title} {Naturalness, chiral symmetry, and
  spontaneous chiral symmetry breaking},\ in\ \href
  {https://doi.org/10.1007/978-1-4684-7571-5_9} {\emph {\bibinfo {booktitle}
  {Recent Developments in Gauge Theories}}},\ \bibinfo {series} {NATO Science
  Series B:}, Vol.~\bibinfo {volume} {59},\ \bibinfo {editor} {edited by\
  \bibinfo {editor} {\bibfnamefont {G.}~\bibnamefont {'t~Hooft}}, \bibinfo
  {editor} {\bibfnamefont {C.}~\bibnamefont {Itzykson}}, \bibinfo {editor}
  {\bibfnamefont {A.}~\bibnamefont {Jaffe}}, \bibinfo {editor} {\bibfnamefont
  {H.}~\bibnamefont {Lehmann}}, \bibinfo {editor} {\bibfnamefont {P.~K.}\
  \bibnamefont {Mitter}}, \bibinfo {editor} {\bibfnamefont {I.~M.}\
  \bibnamefont {Singer}},\ and\ \bibinfo {editor} {\bibfnamefont
  {R.}~\bibnamefont {Stora}}}\ (\bibinfo  {publisher} {Springer {US}},\
  \bibinfo {address} {Boston, MA},\ \bibinfo {year} {1980})\ pp.\ \bibinfo
  {pages} {135--157}\BibitemShut {NoStop}%
\bibitem [{\citenamefont {Zhou}\ \emph {et~al.}(2021)\citenamefont {Zhou},
  \citenamefont {Zhang}, \citenamefont {Pollmann},\ and\ \citenamefont
  {You}}]{Zhou:2021wsv}%
  \BibitemOpen
  \bibfield  {author} {\bibinfo {author} {\bibfnamefont {Z.}~\bibnamefont
  {Zhou}}, \bibinfo {author} {\bibfnamefont {X.-F.}\ \bibnamefont {Zhang}},
  \bibinfo {author} {\bibfnamefont {F.}~\bibnamefont {Pollmann}},\ and\
  \bibinfo {author} {\bibfnamefont {Y.}~\bibnamefont {You}},\ }\bibfield
  {journal} {\bibinfo  {journal} {arXiv e-prints}\ }\href
  {https://doi.org/10.48550/arXiv.2105.05851} {10.48550/arXiv.2105.05851}
  (\bibinfo {year} {2021}),\ \Eprint {https://arxiv.org/abs/2105.05851}
  {arXiv:2105.05851 [cond-mat.str-el]} \BibitemShut {NoStop}%
\bibitem [{\citenamefont {Seiberg}\ and\ \citenamefont
  {Shao}(2020)}]{Seiberg_2020}%
  \BibitemOpen
  \bibfield  {author} {\bibinfo {author} {\bibfnamefont {N.}~\bibnamefont
  {Seiberg}}\ and\ \bibinfo {author} {\bibfnamefont {S.-H.}\ \bibnamefont
  {Shao}},\ }\bibfield  {journal} {\bibinfo  {journal} {SciPost Physics}\
  }\textbf {\bibinfo {volume} {9}},\ \href
  {https://doi.org/10.21468/SciPostPhys.9.4.046} {10.21468/SciPostPhys.9.4.046}
  (\bibinfo {year} {2020}),\ \Eprint {https://arxiv.org/abs/2004.00015}
  {arXiv:2004.00015 [cond-mat.str-el]} \BibitemShut {NoStop}%
\bibitem [{\citenamefont {You}\ \emph {et~al.}(2020{\natexlab{b}})\citenamefont
  {You}, \citenamefont {Devakul}, \citenamefont {Sondhi},\ and\ \citenamefont
  {Burnell}}]{You_2020}%
  \BibitemOpen
  \bibfield  {author} {\bibinfo {author} {\bibfnamefont {Y.}~\bibnamefont
  {You}}, \bibinfo {author} {\bibfnamefont {T.}~\bibnamefont {Devakul}},
  \bibinfo {author} {\bibfnamefont {S.~L.}\ \bibnamefont {Sondhi}},\ and\
  \bibinfo {author} {\bibfnamefont {F.~J.}\ \bibnamefont {Burnell}},\
  }\bibfield  {journal} {\bibinfo  {journal} {Physical Review Research}\
  }\textbf {\bibinfo {volume} {2}},\ \href
  {https://doi.org/10.1103/PhysRevResearch.2.023249}
  {10.1103/PhysRevResearch.2.023249} (\bibinfo {year} {2020}{\natexlab{b}}),\
  \Eprint {https://arxiv.org/abs/1904.11530} {arXiv:1904.11530
  [cond-mat.str-el]} \BibitemShut {NoStop}%
\bibitem [{\citenamefont {Lake}(2022)}]{Lake2021rg}%
  \BibitemOpen
  \bibfield  {author} {\bibinfo {author} {\bibfnamefont {E.}~\bibnamefont
  {Lake}},\ }\bibfield  {journal} {\bibinfo  {journal} {Physical Review B}\
  }\textbf {\bibinfo {volume} {105}},\ \href
  {https://doi.org/10.1103/physrevb.105.075115} {10.1103/physrevb.105.075115}
  (\bibinfo {year} {2022}),\ \Eprint {https://arxiv.org/abs/2110.02986}
  {arXiv:2110.02986 [cond-mat.str-el]} \BibitemShut {NoStop}%
\bibitem [{\citenamefont {Oh}\ \emph {et~al.}(2022)\citenamefont {Oh},
  \citenamefont {Kim}, \citenamefont {Moon},\ and\ \citenamefont
  {Han}}]{Oh_2022}%
  \BibitemOpen
  \bibfield  {author} {\bibinfo {author} {\bibfnamefont {Y.-T.}\ \bibnamefont
  {Oh}}, \bibinfo {author} {\bibfnamefont {J.}~\bibnamefont {Kim}}, \bibinfo
  {author} {\bibfnamefont {E.-G.}\ \bibnamefont {Moon}},\ and\ \bibinfo
  {author} {\bibfnamefont {J.~H.}\ \bibnamefont {Han}},\ }\bibfield  {journal}
  {\bibinfo  {journal} {Physical Review B}\ }\textbf {\bibinfo {volume}
  {105}},\ \href {https://doi.org/10.1103/PhysRevB.105.045128}
  {10.1103/PhysRevB.105.045128} (\bibinfo {year} {2022}),\ \Eprint
  {https://arxiv.org/abs/2110.02658} {arXiv:2110.02658 [cond-mat.str-el]}
  \BibitemShut {NoStop}%
\bibitem [{\citenamefont {Pace}\ and\ \citenamefont {Wen}(2022)}]{Pace_2022}%
  \BibitemOpen
  \bibfield  {author} {\bibinfo {author} {\bibfnamefont {S.~D.}\ \bibnamefont
  {Pace}}\ and\ \bibinfo {author} {\bibfnamefont {X.-G.}\ \bibnamefont {Wen}},\
  }\bibfield  {journal} {\bibinfo  {journal} {Physical Review B}\ }\textbf
  {\bibinfo {volume} {106}},\ \href
  {https://doi.org/10.1103/PhysRevB.106.045145} {10.1103/PhysRevB.106.045145}
  (\bibinfo {year} {2022}),\ \Eprint {https://arxiv.org/abs/2204.07111}
  {arXiv:2204.07111 [cond-mat.str-el]} \BibitemShut {NoStop}%
\bibitem [{\citenamefont {Delfino}\ \emph {et~al.}(2023)\citenamefont
  {Delfino}, \citenamefont {Fontana}, \citenamefont {Gomes},\ and\
  \citenamefont {Chamon}}]{Delfino_2023}%
  \BibitemOpen
  \bibfield  {author} {\bibinfo {author} {\bibfnamefont {G.}~\bibnamefont
  {Delfino}}, \bibinfo {author} {\bibfnamefont {W.~B.}\ \bibnamefont
  {Fontana}}, \bibinfo {author} {\bibfnamefont {P.~R.~S.}\ \bibnamefont
  {Gomes}},\ and\ \bibinfo {author} {\bibfnamefont {C.}~\bibnamefont
  {Chamon}},\ }\bibfield  {journal} {\bibinfo  {journal} {SciPost Physics}\
  }\textbf {\bibinfo {volume} {14}},\ \href
  {https://doi.org/10.21468/SciPostPhys.14.1.002}
  {10.21468/SciPostPhys.14.1.002} (\bibinfo {year} {2023}),\ \Eprint
  {https://arxiv.org/abs/2207.00409} {arXiv:2207.00409 [cond-mat.str-el]}
  \BibitemShut {NoStop}%
\bibitem [{\citenamefont {Watanabe}\ \emph {et~al.}(2023)\citenamefont
  {Watanabe}, \citenamefont {Cheng},\ and\ \citenamefont
  {Fuji}}]{Watanabe_2023}%
  \BibitemOpen
  \bibfield  {author} {\bibinfo {author} {\bibfnamefont {H.}~\bibnamefont
  {Watanabe}}, \bibinfo {author} {\bibfnamefont {M.}~\bibnamefont {Cheng}},\
  and\ \bibinfo {author} {\bibfnamefont {Y.}~\bibnamefont {Fuji}},\ }\bibfield
  {journal} {\bibinfo  {journal} {Journal of Mathematical Physics}\ }\textbf
  {\bibinfo {volume} {64}},\ \href {https://doi.org/10.1063/5.0134010}
  {10.1063/5.0134010} (\bibinfo {year} {2023}),\ \Eprint
  {https://arxiv.org/abs/2211.00299} {arXiv:2211.00299 [cond-mat.other]}
  \BibitemShut {NoStop}%
\bibitem [{\citenamefont {Ma}\ \emph {et~al.}(2017)\citenamefont {Ma},
  \citenamefont {Lake}, \citenamefont {Chen},\ and\ \citenamefont
  {Hermele}}]{Ma_2017}%
  \BibitemOpen
  \bibfield  {author} {\bibinfo {author} {\bibfnamefont {H.}~\bibnamefont
  {Ma}}, \bibinfo {author} {\bibfnamefont {E.}~\bibnamefont {Lake}}, \bibinfo
  {author} {\bibfnamefont {X.}~\bibnamefont {Chen}},\ and\ \bibinfo {author}
  {\bibfnamefont {M.}~\bibnamefont {Hermele}},\ }\bibfield  {journal} {\bibinfo
   {journal} {Physical Review B}\ }\textbf {\bibinfo {volume} {95}},\ \href
  {https://doi.org/10.1103/PhysRevB.95.245126} {10.1103/PhysRevB.95.245126}
  (\bibinfo {year} {2017}),\ \Eprint {https://arxiv.org/abs/1701.00747}
  {arXiv:1701.00747 [cond-mat.str-el]} \BibitemShut {NoStop}%
\bibitem [{\citenamefont {Newman}\ and\ \citenamefont
  {Moore}(1999)}]{Newman_1999}%
  \BibitemOpen
  \bibfield  {author} {\bibinfo {author} {\bibfnamefont {M.~E.~J.}\
  \bibnamefont {Newman}}\ and\ \bibinfo {author} {\bibfnamefont
  {C.}~\bibnamefont {Moore}},\ }\href
  {https://doi.org/10.1103/PhysRevE.60.5068} {\bibfield  {journal} {\bibinfo
  {journal} {Physical Review E}\ }\textbf {\bibinfo {volume} {60}},\ \bibinfo
  {pages} {5068} (\bibinfo {year} {1999})},\ \Eprint
  {https://arxiv.org/abs/cond-mat/9707273} {arXiv:cond-mat/9707273
  [cond-mat.stat-mech]} \BibitemShut {NoStop}%
\bibitem [{\citenamefont {Castelnovo}\ and\ \citenamefont
  {Chamon}(2012)}]{Castelnovo_2012}%
  \BibitemOpen
  \bibfield  {author} {\bibinfo {author} {\bibfnamefont {C.}~\bibnamefont
  {Castelnovo}}\ and\ \bibinfo {author} {\bibfnamefont {C.}~\bibnamefont
  {Chamon}},\ }\href {https://doi.org/10.1080/14786435.2011.609152} {\bibfield
  {journal} {\bibinfo  {journal} {Philosophical Magazine}\ }\textbf {\bibinfo
  {volume} {92}},\ \bibinfo {pages} {304} (\bibinfo {year} {2012})},\ \Eprint
  {https://arxiv.org/abs/1108.2051} {arXiv:1108.2051 [cond-mat.str-el]}
  \BibitemShut {NoStop}%
\bibitem [{\citenamefont {Yoshida}(2013)}]{Yoshida:2013sqa}%
  \BibitemOpen
  \bibfield  {author} {\bibinfo {author} {\bibfnamefont {B.}~\bibnamefont
  {Yoshida}},\ }\href {https://doi.org/10.1103/PhysRevB.88.125122} {\bibfield
  {journal} {\bibinfo  {journal} {Physical Review B}\ }\textbf {\bibinfo
  {volume} {88}},\ \bibinfo {pages} {125122} (\bibinfo {year} {2013})},\
  \Eprint {https://arxiv.org/abs/1302.6248} {arXiv:1302.6248 [cond-mat.str-el]}
  \BibitemShut {NoStop}%
\bibitem [{\citenamefont {Bulmash}\ and\ \citenamefont
  {Barkeshli}(2018)}]{Bulmash:2018knk}%
  \BibitemOpen
  \bibfield  {author} {\bibinfo {author} {\bibfnamefont {D.}~\bibnamefont
  {Bulmash}}\ and\ \bibinfo {author} {\bibfnamefont {M.}~\bibnamefont
  {Barkeshli}},\ }\bibfield  {journal} {\bibinfo  {journal} {arXiv e-prints}\
  }\href {https://doi.org/10.48550/arXiv.1806.01855}
  {10.48550/arXiv.1806.01855} (\bibinfo {year} {2018}),\ \Eprint
  {https://arxiv.org/abs/1806.01855} {arXiv:1806.01855 [cond-mat.str-el]}
  \BibitemShut {NoStop}%
\bibitem [{\citenamefont {Devakul}\ \emph {et~al.}(2019)\citenamefont
  {Devakul}, \citenamefont {You}, \citenamefont {Burnell},\ and\ \citenamefont
  {Sondhi}}]{Devakul_2019}%
  \BibitemOpen
  \bibfield  {author} {\bibinfo {author} {\bibfnamefont {T.}~\bibnamefont
  {Devakul}}, \bibinfo {author} {\bibfnamefont {Y.}~\bibnamefont {You}},
  \bibinfo {author} {\bibfnamefont {F.~J.}\ \bibnamefont {Burnell}},\ and\
  \bibinfo {author} {\bibfnamefont {S.}~\bibnamefont {Sondhi}},\ }\bibfield
  {journal} {\bibinfo  {journal} {{SciPost} Physics}\ }\textbf {\bibinfo
  {volume} {6}},\ \href {https://doi.org/10.21468/SciPostPhys.6.1.007}
  {10.21468/SciPostPhys.6.1.007} (\bibinfo {year} {2019}),\ \Eprint
  {https://arxiv.org/abs/1805.04097} {arXiv:1805.04097 [cond-mat.str-el]}
  \BibitemShut {NoStop}%
\bibitem [{\citenamefont {Myerson-Jain}\ \emph {et~al.}(2022)\citenamefont
  {Myerson-Jain}, \citenamefont {Liu}, \citenamefont {Ji}, \citenamefont {Xu},\
  and\ \citenamefont {Vijay}}]{Myerson_Jain_2022}%
  \BibitemOpen
  \bibfield  {author} {\bibinfo {author} {\bibfnamefont {N.~E.}\ \bibnamefont
  {Myerson-Jain}}, \bibinfo {author} {\bibfnamefont {S.}~\bibnamefont {Liu}},
  \bibinfo {author} {\bibfnamefont {W.}~\bibnamefont {Ji}}, \bibinfo {author}
  {\bibfnamefont {C.}~\bibnamefont {Xu}},\ and\ \bibinfo {author}
  {\bibfnamefont {S.}~\bibnamefont {Vijay}},\ }\bibfield  {journal} {\bibinfo
  {journal} {Physical Review Letters}\ }\textbf {\bibinfo {volume} {128}},\
  \href {https://doi.org/10.1103/PhysRevLett.128.115301}
  {10.1103/PhysRevLett.128.115301} (\bibinfo {year} {2022}),\ \Eprint
  {https://arxiv.org/abs/2110.02237} {arXiv:2110.02237 [cond-mat.str-el]}
  \BibitemShut {NoStop}%
\bibitem [{\citenamefont {Sfairopoulos}\ \emph {et~al.}(2023)\citenamefont
  {Sfairopoulos}, \citenamefont {Causer}, \citenamefont {Mair},\ and\
  \citenamefont {Garrahan}}]{Sfairopoulos2023}%
  \BibitemOpen
  \bibfield  {author} {\bibinfo {author} {\bibfnamefont {K.}~\bibnamefont
  {Sfairopoulos}}, \bibinfo {author} {\bibfnamefont {L.}~\bibnamefont
  {Causer}}, \bibinfo {author} {\bibfnamefont {J.~F.}\ \bibnamefont {Mair}},\
  and\ \bibinfo {author} {\bibfnamefont {J.~P.}\ \bibnamefont {Garrahan}},\
  }\bibfield  {journal} {\bibinfo  {journal} {Physical Review B}\ }\textbf
  {\bibinfo {volume} {108}},\ \href
  {https://doi.org/10.1103/physrevb.108.174107} {10.1103/physrevb.108.174107}
  (\bibinfo {year} {2023}),\ \Eprint {https://arxiv.org/abs/2301.02826}
  {arXiv:2301.02826 [cond-mat.stat-mech]} \BibitemShut {NoStop}%
\bibitem [{Note1()}]{Note1}%
  \BibitemOpen
  \bibinfo {note} {See Supplemental Material \protect \hyperref
  [SM]{Supplementary Material} for further insights and detailed explanations
  related to several topics discussed in the main text. These include the
  determination of \protect \hyperref [MS1]{ground state degeneracy}, the
  \protect \hyperref [MS3]{number of independent Wilson line operators}, how to
  obtain the \protect \hyperref [MS4]{algebra among fractal membranes and
  Wilson line operators} and how\protect \hyperref [MS5]{ Wilson lines can be
  simplified into a product of TD operators} for non-degenerate system
  sizes.}\BibitemShut {Stop}%
\bibitem [{\citenamefont {Kitaev}(2003)}]{Kitaev_2003}%
  \BibitemOpen
  \bibfield  {author} {\bibinfo {author} {\bibfnamefont {A.~Y.}\ \bibnamefont
  {Kitaev}},\ }\href {https://doi.org/10.1016/s0003-4916(02)00018-0} {\bibfield
   {journal} {\bibinfo  {journal} {Annals of Physics}\ }\textbf {\bibinfo
  {volume} {303}},\ \bibinfo {pages} {2} (\bibinfo {year} {2003})},\ \Eprint
  {https://arxiv.org/abs/quant-ph/9707021} {arXiv:quant-ph/9707021 [quant-ph]}
  \BibitemShut {NoStop}%
\bibitem [{Note2()}]{Note2}%
  \BibitemOpen
  \bibinfo {note} {In analogy with the toric code, the two membranes of Eq.
  \protect \eqref {eq:membranes} correspond to contractible loops and a single
  membrane corresponds to a non-contractible loop.}\BibitemShut {Stop}%
\bibitem [{\citenamefont {Pai}\ and\ \citenamefont {Hermele}(2019)}]{Pai2019}%
  \BibitemOpen
  \bibfield  {author} {\bibinfo {author} {\bibfnamefont {S.}~\bibnamefont
  {Pai}}\ and\ \bibinfo {author} {\bibfnamefont {M.}~\bibnamefont {Hermele}},\
  }\bibfield  {journal} {\bibinfo  {journal} {Physical Review B}\ }\textbf
  {\bibinfo {volume} {100}},\ \href
  {https://doi.org/10.1103/PhysRevB.100.195136} {10.1103/PhysRevB.100.195136}
  (\bibinfo {year} {2019}),\ \Eprint {https://arxiv.org/abs/1903.11625}
  {arXiv:1903.11625 [cond-mat.str-el]} \BibitemShut {NoStop}%
\bibitem [{\citenamefont {Song}\ \emph {et~al.}(2023)\citenamefont {Song},
  \citenamefont {Tantivasadakarn}, \citenamefont {Shirley},\ and\ \citenamefont
  {Hermele}}]{Song:2023rml}%
  \BibitemOpen
  \bibfield  {author} {\bibinfo {author} {\bibfnamefont {H.}~\bibnamefont
  {Song}}, \bibinfo {author} {\bibfnamefont {N.}~\bibnamefont
  {Tantivasadakarn}}, \bibinfo {author} {\bibfnamefont {W.}~\bibnamefont
  {Shirley}},\ and\ \bibinfo {author} {\bibfnamefont {M.}~\bibnamefont
  {Hermele}},\ }\bibfield  {journal} {\bibinfo  {journal} {arXiv e-prints}\
  }\href {https://doi.org/10.48550/arXiv.2304.00028}
  {10.48550/arXiv.2304.00028} (\bibinfo {year} {2023}),\ \Eprint
  {https://arxiv.org/abs/2304.00028} {arXiv:2304.00028 [cond-mat.str-el]}
  \BibitemShut {NoStop}%
\bibitem [{Note3()}]{Note3}%
  \BibitemOpen
  \bibinfo {note} {This analysis is also valid for sizes of the form \protect
  \eqref {sizes}. We leave for an upcoming publication a detailed analysis of
  boundary physics.}\BibitemShut {Stop}%
\end{thebibliography}%
